# The Applications of Graph Theory to Investing

Joseph Attia

Brooklyn Technical High School

January 17, 2019

Abstract

How can graph theory be applied to investing in the stock market? The answer may help investors realize the true risks of their investments, help prevent recessions like that of 2008, and increase financial literacy amongst students. Using several original Python programs, we take a correlation matrix with correlations between the stock prices, and then transform that into a graphable binary adjacency matrix. From this graph, we take a graph in which each edge represents weak correlations between two stocks. Finding the largest complete graph will produce a diversified portfolio. Numerous trials have shown that diversified portfolios consistently outperform the market during times of economic stability, but undiversified portfolios prove to be riskier and more unpredictable, either producing huge profits or even larger losses. Furthermore, once deciding among which stocks our portfolio would consist of, how do we know when to invest in each stock to maximize profits? Can taking stock price data and shifting values help predict how a stock will perform today if another stock performs a certain way n days prior? It was found that this method of predicting the optimal time to investment failed to improve returns when based solely on correlations. Although a trial with random stocks with varied correlations produced more profits than continuously investing.

## Table of Contents

# 1. Research Question

How can graph theory be applied to investing and the stock market? Can graphs, in which edges represent correlations, be used to create a diversified or undiversified portfolio that will outperform the index it's based off? In addition, is it possible to use indicators and directed graphs in which edges represent another stock's movement, to dictate when to invest in a specific stock and outperform continuous investing?

# 2. Introduction

"I know what that is", a phrase said confidently by virtually everybody when asked about the stock market. Everyone is familiar with the stock market, yet nobody knows everything about it. A place where the money is endless, adrenaline is high, and risk is even higher.

> *"What counts for most people in investing is not how much they know,*
> *but rather how realistically they define what they don't know."*
>
> *-Warren Buffett*

The quote above by Warren Buffett, one of the most infamous investors of the 21st century, speaks to the fact that in the stock market nobody knows everything about the motion of the market, but those who succeed know what they do not know about the market. As anybody who lacks certain knowledge, how can we learn it, but more importantly how can it benefit us? Is it possible to develop new ways of thinking about investing in the stock market that can help us outperform ordinary investors? That is the goal of my paper. There are numerous different views on the market and different ways to model investing, but is it possible to do so using graph theory? I want to apply basic graph theory to investing and attempt to determine if this modeling helps investors gain the upper hand.

    I chose this topic since I developed an interest in the stock market from a young age. In 6th grade, I got together with a few of my friends and a teacher entered us into the Stock Market Game, a program by the SIGMA Foundation with the intent of "[connecting] students to the global economy with virtual investing and real-world learning." For the first few weeks of the game, we were doing very well, placing in the top 10 teams in the country. Out of nowhere, a week before the contest finished the stocks in our portfolio tanked and our position fell alongside them. I returned to the Stock Market Game in the 9th grade as part of my schools Stock Market Club, and this time, with a different team, we peaked at first place a couple of times but ultimately ended in 5th place nationally. The stock market was exhilarating and fascinating, but more importantly, unpredictable.

    In 11th grade, I began a discrete mathematics course which began with graph theory. After going through Euler graphs and Hamiltonian Graphs and other findings like the four-color theorem, I wondered if I can apply graph theory to better understand the stock market just like in class when we used it to better understand word problems.

— 2 —

# 3. History

## 3.1. Brief History of Graph Theory

Graph Theory can be traced as far back as the early 18th century when Swiss mathematician Leonhard Euler solved the 7 bridges of Königsberg problem. The problem consisted of finding a path that crossed each of the 7 bridges once, and only once. Euler argued that it is impossible.

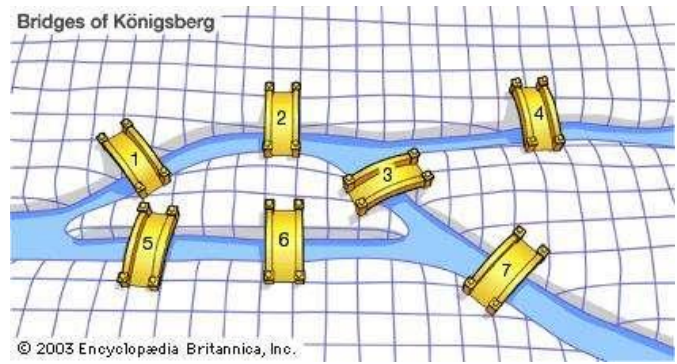

The next major contribution to graph theory came from a mathematician in Ireland, William Rowan Hamilton, who invented a puzzle that involved finding a path where each vertex can only be visited once. This became known as a Hamiltonian Path.

Slowly but surely, this branch of mathematics developed until today when mathematicians and people, in general, have been able to apply it to studies of relationships, software, and more.

## 3.2. History of Mathematical Models

The use of mathematical models, while not as old as graph theory, have been around for a while. The first distinguishable mathematical models were in fact numbers, which can be traced back to 30,000 BC. Following numbers, astronomers and ancient architects began to apply mathematical models. Around the 20th century, the first computer was released and with it a plethora of ways to model math but even better, a way for everyone to develop countless mathematical models themselves.

### 3.2.1. Use of Mathematical Models in Finance

As the personal computer became popular, people began introducing mathematical modeling to finance. One of the best-known mathematical models applied differential equations and Brownian motion, described by Louis Bachelier in 1900 and Albert Einstein in 1905, in order to estimate the price of a European option. This model is called the Black-Scholes model, was developed in 1973, and was heavily used throughout the following decades.

### 3.3. History of the Stock Market

The first stock exchange was established in Belgium in 1531 where brokers and moneylenders met in order to create deals with businesses, governments, and even individual debt. In the 17th century, charters were granted to companies in many different European countries which granted governments a stake in the profits in the East. Nevertheless, many ship owners began seeking more investors so that if anything happened to their ships while at sea their fortune would not be ruined. Investors would also manage their risk by investing in many different ventures, ensuring profits would cover their losses.

The London Stock Exchange was opened in 1773 and the New York Stock Exchange (NYSE) was opened 19 years after. Unlike the London Stock Exchange, who was restricted from selling shared, the NYSE sold stocks from its birth. The NYSE was located in one of the most economically thriving cities in America on Wall Street, it quickly became the most popular stock exchange in the country.

In 1971, nearly 200 years after the inception of the NYSE, the Nasdaq was created. Developed by the Financial Industry Regulatory Authority, the Nasdaq was the first of its kind in the world since it didn't take up a physical building. It was a network of computers that executed all trades electronically, which in turn made trading more efficient. This competition forced the NYSE to step up its game by merging with a European stock exchange and becoming the first worldwide exchange.

## 4. Mathematical Background

### 4.1. Intro to Graph Theory

A graph **G** consists of pairs of sets **(V,E)** where **V** represents vertices/nodes and **E** represents edges that connect pairs of vertices. Here's an example:

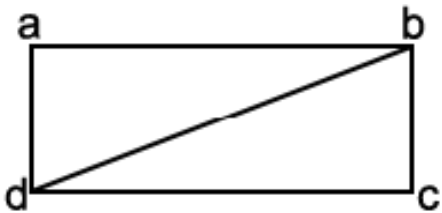

V = {a, b, c, d}

E = {ab, bc, bd, cd, da}

When two points are connected with an edge, they can also be classified as adjacent points. Therefore, every simple graph (only one possible connection between ever two points) can also be represented by an adjacency matrix that represents the graph and connections between vertices. The preceding graph can be shown as the following adjacency matrix.

$$\begin{array}{c} \phantom{A} \quad A \ \ B \ \ C \ \ D \\ \begin{array}{c} A \\ B \\ C \\ D \end{array} \begin{bmatrix} [0. & 1. & 0. & 1.] \\ [1. & 0. & 1. & 1.] \\ [0. & 1. & 0. & 1.] \\ [1. & 1. & 1. & 0.] \end{bmatrix} \end{array}$$

An adjacency matrix for simple graphs contains 0's and 1's that represent either an edge or an absence of an edge. The diagonal line that starts at the top left and continues to the bottom right, always contains 0's since no vertex can be connected to itself. This graph is also undirected, so the matrix will be symmetrical over the diagonal line of 0's previously described. The matrix above shows that every edge exists except for an edge between vertices c and a.

When a graph contains all possible edges, an edge connecting every possible pair of points, the adjacency matrix contains all 1's other than the diagonal line. This type of graph is called a complete graph. Here's an example of the first few complete graphs.

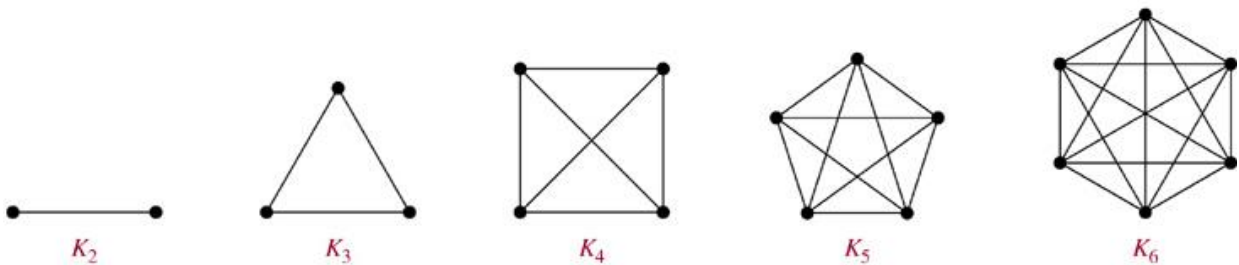

The graphs above all have undirected edges, which means that if vertices a and b are connected, whatever the edge represents (in this paper an edge signifies correlation) can be applied from both a to b, and vice-a-versa. There are also directed graphs that signify the relationship only goes one way. These can be represented in the same way as matrices except it won't be symmetrical and depending on whether the 1 is on the row or column, it would show the direction.

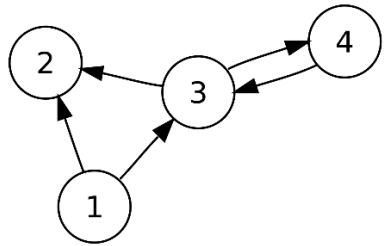

## 4.2. Intro to Investing and Stocks

When analyzing historical stock data, analysts have a variety of different options for which data they would like to use. Ways to represent stock prices include open price (price at beginning of the trading day), close price (the price before the trading day ends), high (the highest point of the stock during the trading day), low (the lowest point of the stock during the trading day), change (the change in stock price, which can be in dollars or percent), volume (the total amount of shares traded for the trading day), adjusted closing price (a closing price that has been adjusted to account for any splits and dividends).

| Date | Open | High | Low | Close | Adj Close | Volume |
|---|---|---|---|---|---|---|
| 2017/12/1 | null | null | null | null | null | null |
| 2018/1/1 | 1187.32 | 1273.99 | 1187.32 | 1251.42 | 1251.420044 | 73366640000 |
| 2018/2/1 | 1248.27 | 1258.88 | 1120.08 | 1201.87 | 1201.869995 | 79579410000 |
| 2018/3/1 | 1202.46 | 1235.97 | 1131.98 | 1157.37 | 1157.369995 | 76349800000 |

The data used in this paper will all be adjusted closing price because it removes the need to adjust stock pricing for splits and dividends, thus making the analysis of the data accurate.

A split is when a company decides to split each share into multiple shares, each at a lower value, thus making the stock more marketable. For example, if one share of stock Z is worth $300. The company decides to split the stock one hundred-for-one. Now the company has 100 times more shares available to trade, each at a price of $3. Nevertheless, if an investor owned 1 share before the split, he would own 100 shares after the split and therefore maintains the value of his investment.

A dividend is when a company distributes part of its profits back to its investors, which in turn lowers their stock price by the same amount. For example, stock B declares a $5 cash dividend and is trading at $105 dollars per share before the dividend date. On the dividend date, the stock price is reduced by $5, and the adjusted closing price becomes $100.

Another term commonly used is diversification. Diversification is a risk management technique that strives to have a collection of stocks among which a negative move in one will be negated by the positive move of the other. Usually, this is achieved by taking stocks from different sectors (financials, utilities, energy, industrials, technology, telecom, materials, real estate, etc.) with the logic that an event that effects technology is unlikely to affect real estate in the same way. In this paper, however, in order to diversify we will be looking at correlations. A diversified portfolio consists of stocks that are very loosely correlated with each other (correlation closer to 0) and a diversified portfolio consists of stocks that are more correlated with each other (correlation closer to 1 or -1)

Like graphs, correlations can also be represented in a matrix. This is called a correlation matrix and the cross-section of column A and row C is the correlation between stock A and C.

$$\begin{array}{cc} & \begin{array}{cc} A & B \end{array} \\ \begin{array}{c} A \\ B \end{array} & \left[\begin{array}{cc} [1. & 0.3\,] \\ [0.3 & 1.\phantom{0}\,] \end{array}\right] \end{array}$$

The correlation matrix above shows the correlation between stock A and stock B. Stock A has a perfect correlation with itself and therefore has a value of 1. The same for Stock B. Stock B and stock A have the same correlation as Stock A and stock B and therefore have the same correlation value of 0.3. The formula for the Pearson method of correlation (Pearson Product Moment Correlation [PPMC]) is

$$Correlation(X,Y) = \frac{\Sigma(x-\bar{x})(y-\bar{y})}{\sqrt{\Sigma(x-\bar{x})^2 \Sigma(y-\bar{y})^2}}$$

The correlation formula outputs the linear relationship between two sets of data **X** and **Y**. Let's look at some examples.

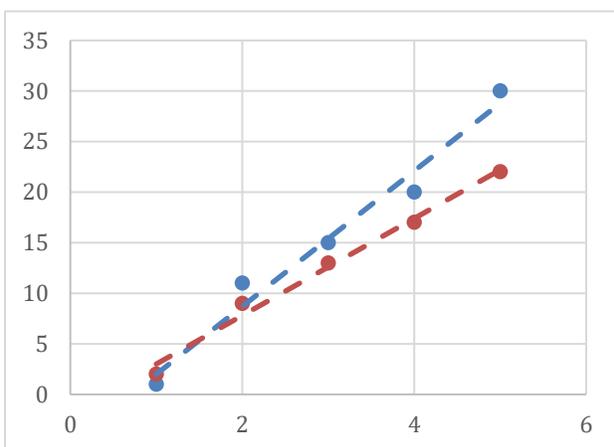

| X | Y₁ | Y₂ |
|---|----|----|
| 1 | 1  | 2  |
| 2 | 11 | 9  |
| 3 | 15 | 13 |
| 4 | 20 | 17 |
| 5 | 30 | 22 |

These points look positively correlated and are seemingly moving together. Let's check this with the correlation formula which should produce a result close to 1.

$$Correlation(Y_1, Y_2) = \frac{[(1-15.4)(2-12.6)][(11-15.4)(9-12.6)]\ldots}{\sqrt{[(1-15.4)^2 + (11-15.4)^2 \ldots][(2-12.6)^2 + (9-12.6)^2 \ldots]}}$$

$$Correlation(Y_1, Y_2) = \frac{320.5}{\sqrt{[233.2][447]}}$$

$$Correlation(Y_1, Y_2) = 0.9927$$

A second example to show what a weak negative correlation signifies.

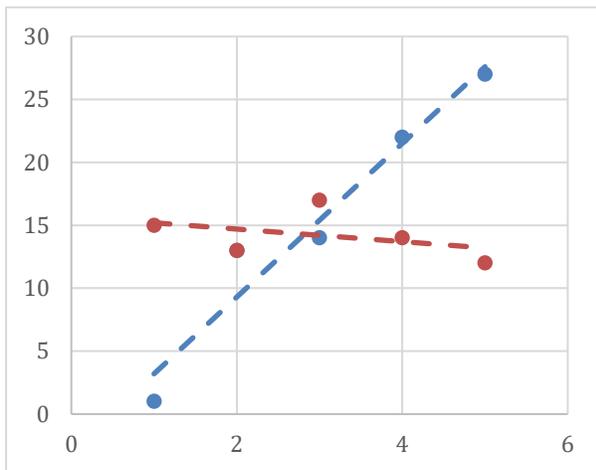

| X | $Y_1$ | $Y_2$ |
|---|---|---|
| 1 | 1 | 15 |
| 2 | 13 | 13 |
| 3 | 14 | 17 |
| 4 | 22 | 14 |
| 5 | 27 | 12 |

These points now look negatively correlated and are moving the opposite way from each other generally but very strongly since there are exceptions. Let's check this with the correlation formula which should produce a result close to 1.

$$Correlation(Y_1, Y_2) = \frac{[(1-16.6)(15-14.2)][(13-16.6)(13-14.2)]\ldots}{\sqrt{[(1-16.6)^2 + (11-16.6)^2 \ldots][(15-14.2)^2 + (13-14.2)^2 \ldots]}}$$

$$Correlation(Y_1, Y_2) = \frac{-22.6}{\sqrt{[405.2][14.8]}}$$

$$Correlation(Y_1, Y_2) = -0.2918$$

## 4.3. Statistical Analysis

In our research, we will be finding the correlations between every pair of stocks in a given set and creating a correlation matrix. Then, we will use an input threshold to change those values to either ones or zeroes based on if it is greater or less than the threshold value. The resulting matrix is an adjacency matrix in which nodes represent stocks and edges represent the lowest or highest correlations.

<u>Correlation Matrix</u>                                          <u>Adjacency Matrix</u>

```
[[0.  0.2 0.4 0.4 0.1]          Diversified &          [[0. 1. 0. 0. 1.]
 [0.2 0.  0.2 0.3 0.2]        a threshold of 0.21       [1. 0. 1. 0. 1.]
 [0.4 0.2 0.  0.6 0.1]         ───────────────▶         [0. 1. 0. 0. 1.]
 [0.4 0.3 0.6 0.  0.1]                                  [0. 0. 0. 0. 1.]
 [0.1 0.2 0.1 0.1 0. ]]                                 [1. 1. 1. 1. 0.]]
```

When analyzing correlations between pairs of dozens of stocks, it may be hard to develop thresholds and correctly analyze the graph. To better assist the user to determine thresholds it is useful to analyze the data sets and provide certain statistics about the data set.

The mean of a data set is a measure of central tendency and is discovered by dividing the sum of all the values by the number of values. Mean fails to account for outliers.

$$mean\ (\bar{x}) = \frac{1 + 2 + 3 + 4 + 6 + 18 + 22 + 92 + 100 + 201 + 300}{11} = 68\frac{1}{11}$$

The median of a data set also measures central tendency but is calculated by putting all numbers in numeric order and removing values one at a time until a central one is found, or the mean of the two remaining values if there is an even amount of values.

Median fails to account for clustering of lower numbers even though the other values may be far away.

$$1\ 2\ 3\ 4\ 6\ \boxed{18}\ 22\ 92\ \cancel{100}\ \cancel{201}\ \cancel{300}$$

When suggesting a threshold to the user, how do we determine which threshold will provide the right amount of edges? To do this we will use standard deviations in order to make sure to get a certain percentage of the edges in our graph. The mean of the dataset minus one standard deviation provides a graph that retains edges between 16% of the lest correlated stocks. A mean plus one standard deviation gives a graph that shows connections between the 16% most correlated stocks.

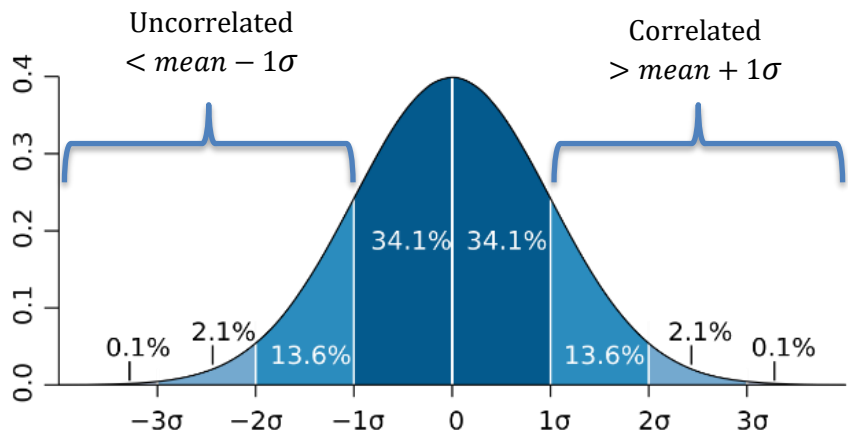

The standard deviation formula is based on the average deviation, the average of the distance from each data value to the mean.

$$average\ deviation = \frac{|x_1 - \bar{x}| + |x_2 - \bar{x}| + \cdots + |x_n - \bar{x}|}{n}$$

In order to eliminate absolute values which, become difficult to work with larger numbers and datasets, let's square the numerator. This provides us with the variance formula which measures the squared deviation. To make this deviation just square root it.

$$variance = \frac{(x_1 - \bar{x})^2 + (x_1 - \bar{x})^2 + \cdots + (x_1 - \bar{x})^2}{n}$$

$$= \sqrt{\frac{(x_1 - \bar{x})^2 + (x_1 - \bar{x})^2 + \cdots + (x_1 - \bar{x})^2}{n}}$$

This formula can be rewritten as a summation function instead of the numerator.

$$= \sqrt{\frac{\Sigma(x - \bar{x})^2}{n}}$$

Now if you have ever seen the standard deviation formula you may ask why the denominator above is n and not n-1. This is what is called the degrees of freedom which is meant to show that if you do not have a final value in the dataset it is possible to find it. If we know the mean of the dataset beforehand you only need n-1 data points and the final value can be figured out.

$$\sigma = \sqrt{\frac{\Sigma(x - \bar{x})^2}{n - 1}}$$

# 5. Investigation

## 5.1. To Diversify or not to Diversify?

### 5.1.1. Program Development

The first part of my investigation will attempt to determine whether diversified portfolios are more return effective than undiversified portfolios. To do this we'll have to write a python program that will create a graph of input stocks based on the correlations between them. I began working on the program and along the way required the following libraries.

```python
8.  from pandas_datareader import data as pdr
9.  import pandas as pd
10. import numpy as np
11. import fix_yahoo_finance as yf
12. import matplotlib.pyplot as plt
13. import statistics as statistics
14. import networkx as nx
```

Then we moved onto the inputs of tickers into an array of strings as well as dates for the beginning and end of the data set.

```python
18. tickers = []
19.
20. ticker_input = input("Please enter the stock tickers you would like to use one by one. When you are done just type 'DONE'! \n")
21.
22. while ticker_input != "DONE":
23.     tickers.append(str(ticker_input))
24.     ticker_input = input()
25.
26. start_date = input("Here enter the date from which data will begin to be taken from (Please put it into the format of YYYY-MM-DD)\n")
27.
28. end_date = input("Here enter the date at which we will stop taking data from (Please put it into the format of YYYY-MM-DD)\n")
29.
30. #Downloads data from Yahoo
31. data = pdr.DataReader(tickers, 'yahoo', start_date, end_date)['Adj Close']
```

Observing resulting data made it clear that comparing price changes in dollars isn't accurate because it does not consider starting price. For example, Stock A opens at $0.01 and closes at $0.02, and Stock B opens at $1000 and closes at $1001. Even though Stock B moved 100 times more in dollars, Stock A doubled in price whereas Stock B only increased by a tenth of a percent. Investing $1000 into Stock A would turn into $2000 dollars, whereas investing $1000 dollars into Stock B would become only $1001. Pandas had a built-in function to change to percent change. The formula for percent change is below, and since for it you need the value before it, the first value in our set becomes an error, so we remove it using a built-in function in pandas.

$$Percent\ Change = \frac{original - new}{original}$$

```
33. #Changes values to percent change for uniform measurements
34. data = data.pct_change()
35.
36. #Removes first percent change and missing data as its NaN
37. data = data.dropna(axis=0, how='any', thresh=None, subset=None, inplace=False)
```

Now we can find the correlations between the datasets using the formula!

$$Correlation(X,Y) = \frac{\sum (x - \bar{x})(y - \bar{y})}{\sqrt{\sum (x - \bar{x})^2 \sum (y - \bar{y})^2}}$$

```
45. #Defines function correlation that can be used to find correlation between two different data sets
    for different stocks
46. def corr(x,y):
47.     x_avg = x.mean()
48.     y_avg = y.mean()
49.     numerator_sum = 0
50.     denominator_sum_x = 0
51.     denominator_sum_y = 0
52.     for i in range(len(x)):
53.         numerator_sum += (x[i] - x_avg)*(y[i] - y_avg)
54.         denominator_sum_x += (x[i] - x_avg) ** 2
55.         denominator_sum_y += (y[i] - y_avg) ** 2
56.     correlation_value = (numerator_sum / ((denominator_sum_x*denominator_sum_y) ** (0.5)))
57.     return correlation_value
58.
59.
60. correlations = np.zeros((len(tickers),len(tickers)))
61. correlations_list = []
62.
63. #Makes correlation matrix for all inputted stock tickers at beginning of program
64. for ticker in range(0,len(tickers)):
65.     for tickerpair in range(0,len(tickers)):
66.         if ticker != tickerpair:
67.             correlations[ticker,tickerpair] = corr(data[:,ticker],data[:,tickerpair])
68.             correlations_list.append(corr(data[:,ticker],data[:,tickerpair]))
```

Now that we know our correlation function works. Let's double check that program also works with a larger input. An input of AAPL (Apple), GOOG (Alphabet), CX (Cemex), FB (Facebook), and T (AT&T) produced the following correlation matrix as a result.

$$\begin{bmatrix} 0. & 0.29776174 & 0.41549919 & 0.48168096 & 0.11685589 \\ 0.29776174 & 0. & 0.28608064 & 0.3142884 & 0.20039596 \\ 0.41549919 & 0.28608064 & 0. & 0.65710632 & 0.06741857 \\ 0.48168096 & 0.3142884 & 0.65710632 & 0. & 0.11252147 \\ 0.11685589 & 0.20039596 & 0.06741857 & 0.11252147 & 0. \end{bmatrix}$$

This same matrix is already formatted correctly to create a graph with nodes and edges, however, the values must be changed. The following snippet of code compares the correlation values to a user chosen threshold and based on that changes the value to a 1, to signify an edge, or a 0, to signify the absence of an edge.

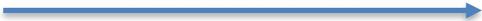

Correlation Matrix

[[0.  0.2 0.4 0.4 0.1 ]
 [0.2 0.  0.2 0.3 0.2 ]
 [0.4 0.2 0.  0.6 0.1 ]
 [0.4 0.3 0.6 0.  0.1 ]
 [0.1 0.2 0.1 0.1 0.  ]]

Diversified & a threshold of 0.21

Adjacency Matrix

[[0. 1. 0. 0. 1. ]
 [1. 0. 1. 0. 1. ]
 [0. 1. 0. 0. 1. ]
 [0. 0. 0. 0. 1. ]
 [1. 1. 1. 1. 0. ]]

```
104. #Makes the previous correlation matrix an adjacency matrix for stocks an takes into account the correlation and threshold.
105. while threshold != "DONE":
106.     for ticker in range(0,len(tickers)):
107.         for tickerpair in range(0,len(tickers)):
108.             if ticker != tickerpair:
109.                 if user_preference == "D":
110.                     if abs(corr(data[:,ticker],data[:,tickerpair])) < threshold:
111.                         adjacency_matrix[ticker,tickerpair] = 1
112.                     if abs(corr(data[:,ticker],data[:,tickerpair])) > threshold:
113.                         adjacency_matrix[ticker,tickerpair] = 0
114.                 if user_preference == "U":
115.                     if abs(corr(data[:,ticker],data[:,tickerpair])) > threshold:
116.                         adjacency_matrix[ticker,tickerpair] = 1
117.                     if abs(corr(data[:,ticker],data[:,tickerpair])) < threshold:
118.                         adjacency_matrix[ticker,tickerpair] = 0
```

If the user seeks an uncorrelated portfolio and indicates it by typing in 'D' when asked, values under the threshold are changed to a 1, and if the user seeks a correlated portfolio and had previously typed in 'U', values over the threshold are set to 1.

Finding largest complete subgraphs by eye is very difficult since there are many more edges than those in the one subgraph we are searching for and the nodes aren't in a polygonal shape, so we need a function that would help highlight the largest complete subgraph for the user. Luckily networkx has a built-in function for complete subgraphs or cliques that will list all complete subgraphs in a larger graph in order from smallest to largest. We can choose the last item in the list of complete subgraphs and before graphing the whole graph, change the color of the tickers in it to a different color.

```
129. complete_graphs = [s for s in nx.enumerate_all_cliques(H) if len(s) > 1]
130.     max_complete_graph = complete_graphs[len(complete_graphs)-1]
131.     print(max_complete_graph)
132.
133.     color_map = []
134.     for index in range (0,len(tickers)):
135.         if tickers[index] in max_complete_graph:
136.             color_map.append('#C21807')
137.         if tickers[index] not in max_complete_graph:
138.             color_map.append('black')
139.
140. nx.draw(H,node_color=color_map,with_labels=True,node_size=450,font_size=7,font_color="white")
```

After testing with various sets of tickers, I realized that finding the perfect threshold was difficult to do with no info on the data, so I decided to provide the means and medians to the user to assist them.

```
72. #Prints the average correlation for the user
73. print("\nThe average correlation is",statistics.mean(correlations_list))
74.
75. #Prints the average correlation for the user
76. print("\nThe median correlation is",statistics.median(correlations_list))
```

I tested it a couple more times and I saw that it was still difficult to determine the optimal threshold. From here I decided to use standard correlations to make sure to retain the 16% lowest correlations (if the user is searching for diversified/uncorrelated stocks), or 16% highest correlations (if the user is searching for undiversified/correlated stocks) are represented on the graph.

```
82. #Finds and utilizes standard deviation to suggest a threshold to the user for the given dataset
83. for i in range(0,len(correlations_list)):
84.     standard_deviation_sum += (correlations_list[i] - statistics.mean(correlations_list)) ** 2
85. standard_deviation = ((standard_deviation_sum / (len(correlations_list) - 1)) ** (0.5))
```

Even after using statistics and standard deviations I found that it was still producing results that were unlikely to create the ideal portfolio the first time around.

Due to this need for multiple trials to find the optimal threshold, I decided to make an interactive graph. It would provide the user with means and medians with the data as well as a suggested threshold using standard deviations, just like before, and the user would enter a value and the graph would show. However, if the threshold didn't produce the desired results, they can close the window and input a new threshold value without having to reenter all previous inputs.

The program to choose the tickers your portfolio will consist of is complete. The next program will serve the purpose of emulating the price of the developed portfolio and compare it to another stock in order to gauge the strategies effectiveness.

I started with the same input system as the first program. After the program continues to calculate the starting price of the portfolio by adding up the prices of one share of each stock in the portfolio. To create accurate results, portfolios must be weighted in the same way as the indexes they are being compared to.

The Dow Jones is weighted using the price-weighted method which consists of ranking companies based on their share price. This method does account for stock splits and we avoid having to do this by using adjusted closing price. This method also does not account for the fact that a $1 change for a $10 stock is much more significant than a $1 change for a $100 stock.

The S&P 500 is weighed using market-capitalization in which companies are ranked by the number of outstanding shares the index holds. To emulate this, an associative list is created that defines a ticker with the number of shares of that stock that is held by the S&P 500.

```python
139. #Function that calculates the starting price of one share of the portfolio given the row number.
140. def portfolio_calc(row):
141.     price = 0
142.     if index_input == "SPY":
143.         for column in range (0,len(tickers)):
144.             price += (data[tickers[column]].iloc[row]*WEIGHT_DICTIONARY[tickers[column]])
145.             sum_of_shares += WEIGHT_DICTIONARY[tickers[column]]
146.
147.     if index_input == "^DJI":
148.         price = data.sum(axis=1)[row]/len(tickers)
149.     if index_input != "^DJI" and "SPY":
150.         price = data.sum(axis=1)[row]
151. return price
```

The user is asked to input the index the portfolio will be compared to. A new data frame is created in which total portfolio price is one column and index price is the other column for the same dates.

```python
136. index_input = str(input("What Index would you like to compare your portfolio to?\n"))
137. index_data = pdr.DataReader(index_input, 'yahoo', start_date, end_date)['Adj Close']
156. index_list = []
157. portfolio_list = []
158.
159. for r in range (0,len(data.index)):
160.     index_list.append (index_data.iloc[r])
161.     portfolio_list.append(portfolio_calc(r))
162.
163. comparison = pd.DataFrame(index=data.index)
164. comparison['Index Price'] = index_list
165. comparison['Portfolio Price'] = portfolio_list
```

The resultant data frame once again has data in dollars, so we must change it to percent change again in order to accurately compare the movement of our portfolio compared to the chosen index. We create a function that if given column name and row will find percent change and output it to a totally new data frame.

```python
156. #Makes a new DataFrame that changes dollar price to percent change from the value prior.
157. comparison_percentage = pd.DataFrame( index = data.index)
158.
159.
```

```
160. def percent_change(Column,row):
161.     if Column == 1:
162.         percentage = ((comparison['Index Price'].iloc[r] - comparison['Index Price'].iloc[0])
163.                         / comparison['Index Price'].iloc[0])
164.     if Column == 2:
165.         percentage = ((comparison['Portfolio Price'].iloc[r] -
    comparison['Portfolio Price'].iloc[0]) / comparison['Portfolio Price'].iloc[0])
166.     return percentage
167.
168.
169. index_list_percent = []
170. portfolio_price_percent = []
171.
172. for r in range (0,len(comparison_percentage.index)):
173.     index_list_percent.append(percent_change(1,r))
174.     portfolio_price_percent.append(percent_change(2,r))
175.
176. comparison_percentage['Index Price'] = index_list_percent
177. comparison_percentage['Portfolio Price'] = portfolio_price_percent
```

Now the graph that is shown accurately compares the movement of our portfolio and the index. As a finishing touch, I added a snippet of code that counts the percentage of days in which the user's portfolio outperforms the index.

```
191. outperformance_count = 0
192. total_count = 0
193.
194. for r in range (0, len(comparison_percentage.index)):
195.     if comparison_percentage['Portfolio Price'].iloc[r] > comparison_percentage['Index Price'].ilo
    c[r]:
196.         outperformance_count += 1
197.     total_count += 1
198. outperformance_percentage = (outperformance_count / total_count) * 100
199.
200. print(comparison_percentage)
201. print("Your portfolio, which consists of", tickers, ", outperforms", index_input, outperformance_p
    ercentage,"% of the time.")
```

### 5.1.2. Testing Effectiveness

To test the effectiveness of our strategy and the success of the programs developed in the following section, we will be taking two of the most well-known indices of the stock market (S&P 500 [SPY] and Dow Jones Index [^DJI]) and attempting to create a portfolio that both outperforms the index and determines whether or not diversified or undiversified portfolios are more efficient.

Since the Dow Jones Index only consist of 30 stocks, it is possible to work with all of them in our testing, although with the S&P 500, working with all 500 stocks may not be very time-efficient. We will be testing with the 50 most weighted stocks within the index.

Beginning with the Dow Jones Index which consists of the following 30 stocks.

| GE | XOM | PG | UTX | MMM | IBM | MRK | AXP | MCD | BA |
| KO | CAT | DIS | JPM | JNJ | HD | INTC | MSFT | PFE | VZ |
| CVX | CSCO | TRV | UNH | GS | NKE | V | AAPL | DWDP | WMT |

Inputting those tickers into our program, taking data from January 1st of 2010 to 2014, and finding a threshold (0.46) for a diversified portfolio produced the following graph.

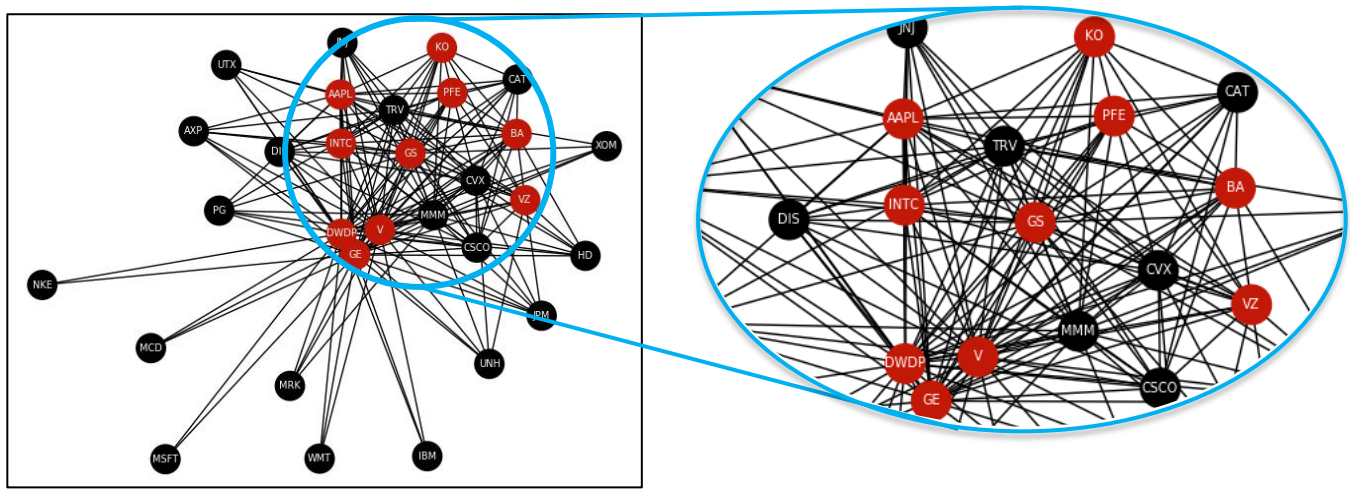

The largest complete subgraph consists of 10 out of the original 30 stocks. They are General Electric (GE), Boeing (BA), Coca-Cola (KO), Intel (INTC), Pfizer (PFE), Verizon (VZ), Goldman Sachs (GS), Visa (V), Apple (AAPL), and DowDuPont (DWDP).

When emulating the portfolio, it is visible both from the graph and the calculated percentage, that the diversified portfolio greatly outperforms the index it is based off.

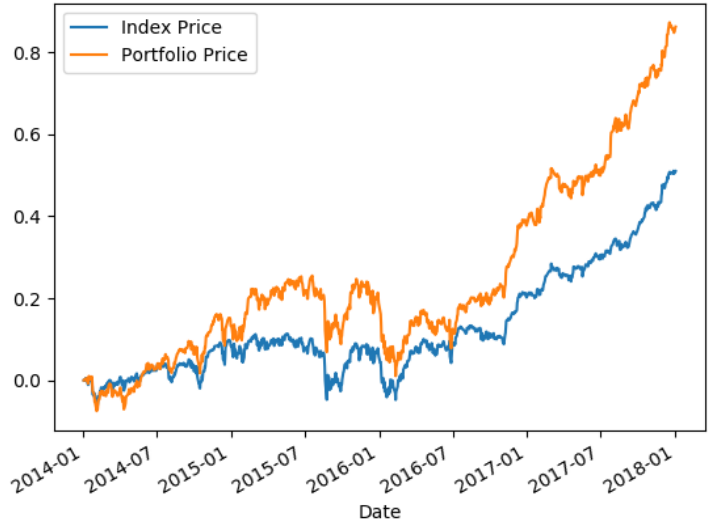

Your diversified portfolio, which consists of ['GE', 'BA', 'KO', 'INTC', 'PFE', 'VZ', 'GS', 'V', 'AAPL', 'DWDP'], outperforms ^DJI 91.17 % of the time.

Now we do the same for an undiversified portfolio. The following graph is the output using the same base tickers, and the same dates but this time with a threshold of 0.6.

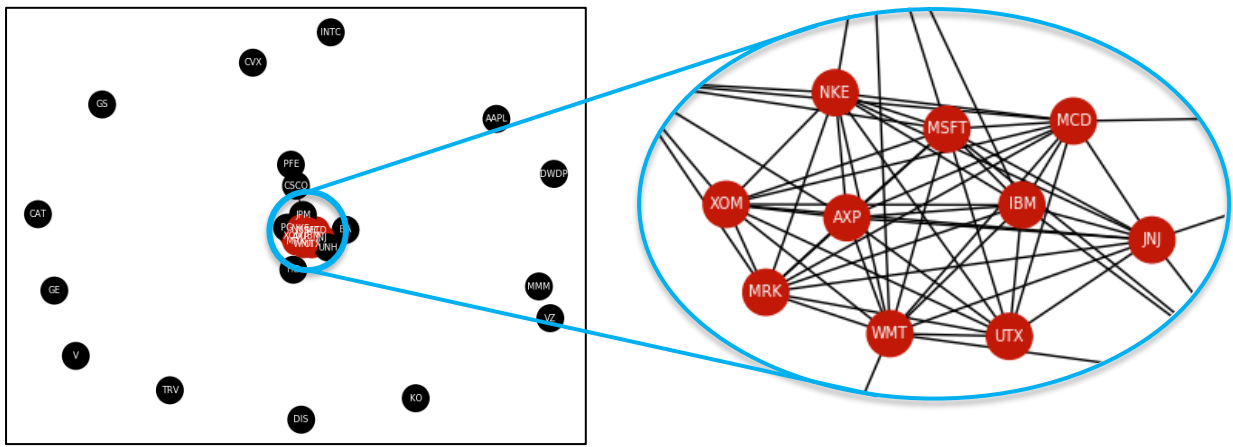

The subgraph that most closely resembles a complete graph consists of 10 out of the original 30 stocks. They are ExxonMobil (XOM), United Technologies (UTX), International Business Machines (IBM), Merck (MRK), American Express (AXP), McDonald's (MCD), Johnson & Johnson (JNJ), Microsoft (MSFT), Nike (NKE), and Walmart (WMT). From there entering it into the portfolio emulation program, it is visible both from the graph and the percent of times it occurs, that the diversified portfolio falls behind the Dow Jones Index.

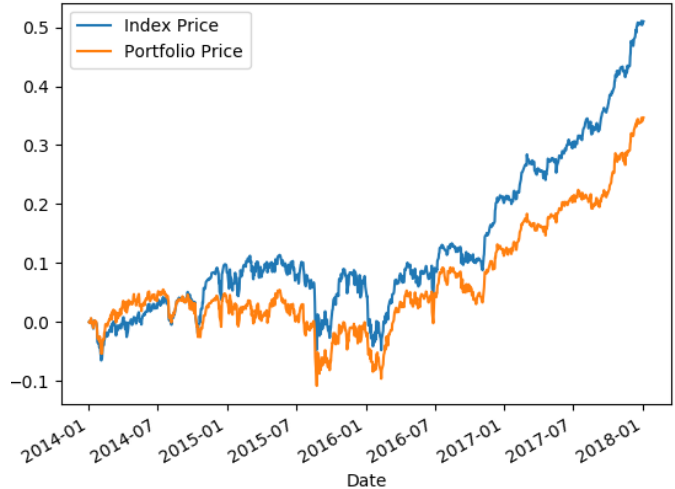

Your undiversified portfolio, which consists of ['XOM', 'UTX', 'IBM', 'MRK', 'AXP', 'MCD', 'JNJ', 'MSFT', 'NKE', 'WMT'] , outperforms ^DJI 20.44 % of the time.

From this data, it's possible to see that a diversified portfolio outperforms the index, but the undiversified portfolio falls behind it. How would they both compare in times of economic recession? To test this, we will utilize data from before the most recent major economic recession, the Great Recession or the 2008 financial crisis. To test this, we will use data from January 1st of 2005 to 2008 (all stocks but Visa (V) have historical data dating this far back) and emulating it for January 1st, 2008 to 2010.

An analysis of the least correlated stocks using data from January 1st, 2005 to 2008 and a threshold of 0.33 resulted in this graph.

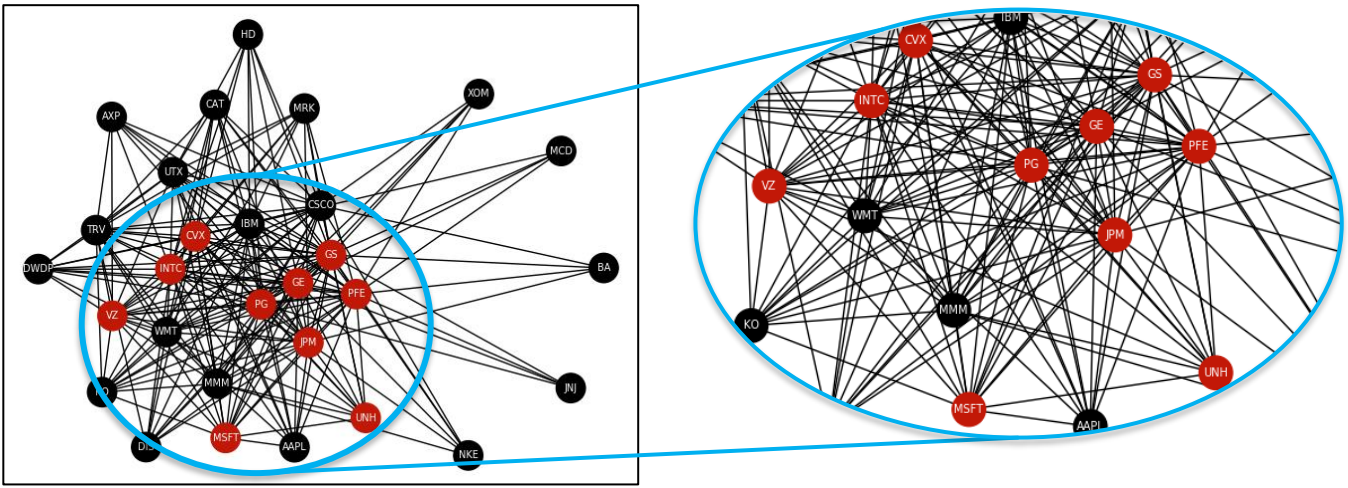

The subgraph that most closely resembles a complete graph consists of 10 out of the original 29 stocks. They are General Electric (GE), Proctor & Gamble (PG), JP Morgan Chase (JPM), Intel (INTC), Microsoft (MSFT), Pfizer (PFE), Verizon (VZ), American Express (CVX), UnitedHealth (UNH), and Goldman Sachs (GS). The portfolio repeatedly falls behind the index during the Great Recession and only recuperates after the economy stabilizes.

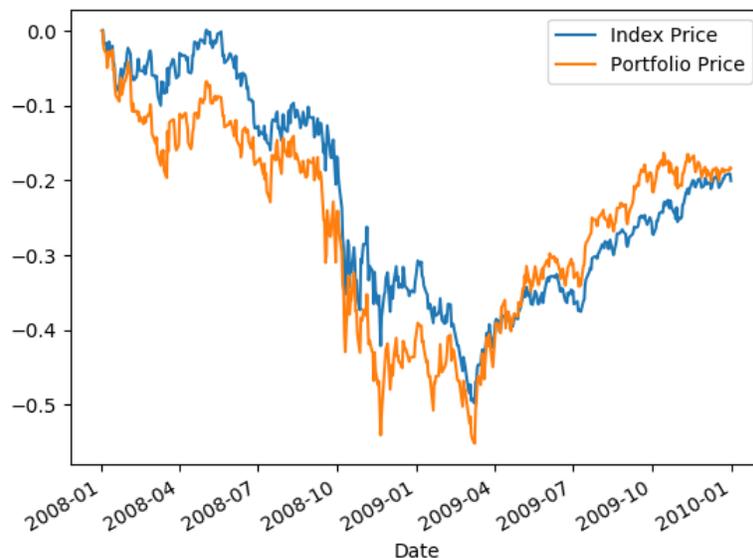

Your diversified portfolio, which consists of ['GE', 'PG', 'JPM', 'INTC', 'MSFT', 'PFE', 'VZ', 'CVX', 'UNH', 'GS'] , outperforms ^DJI 36.44 % of the time.

Attempting the same with an undiversified consisting of the most correlated stocks of the original 29 produces this graph when a threshold of 0.3645 is used.

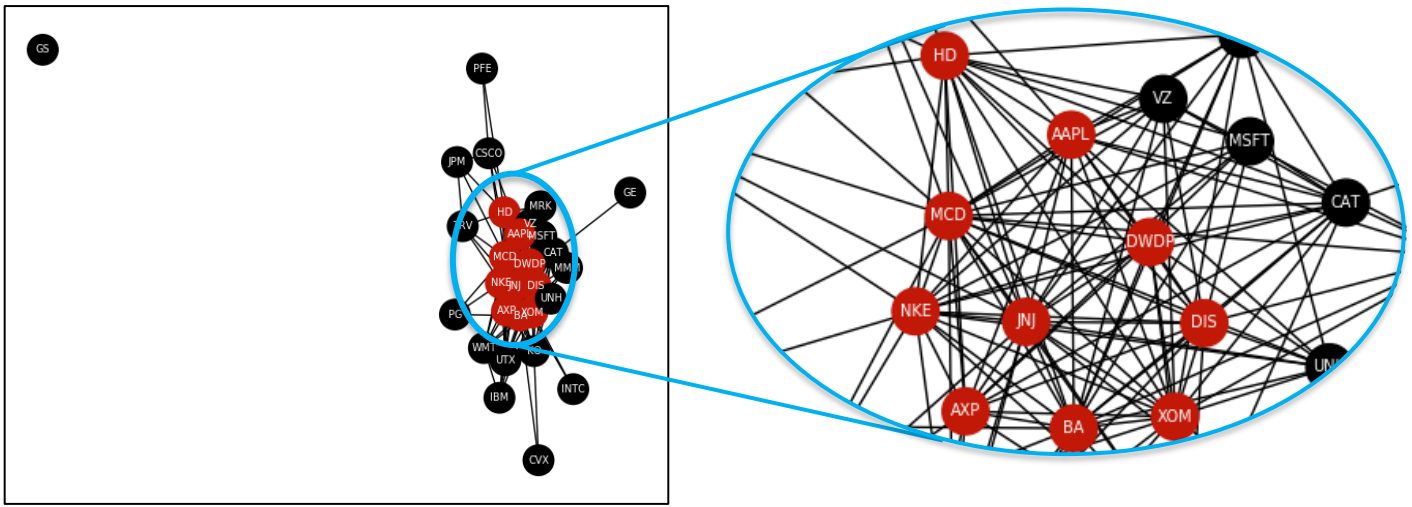

The subgraph that most closely resembles a complete graph consists of 10 out of the original 29 stocks. They are ExxonMobil (XOM), American Express (AXP), McDonald's (MCD), Boeing (BA), Disney (DIS), Johnson & Johnson (JNJ), Home Depot (HD), Nike (NKE), Apple (AAPL), and DowDuPont (DWDP). The portfolio, unlike the diversified one, outperforms the index and reports less of a loss during the Great Recession.

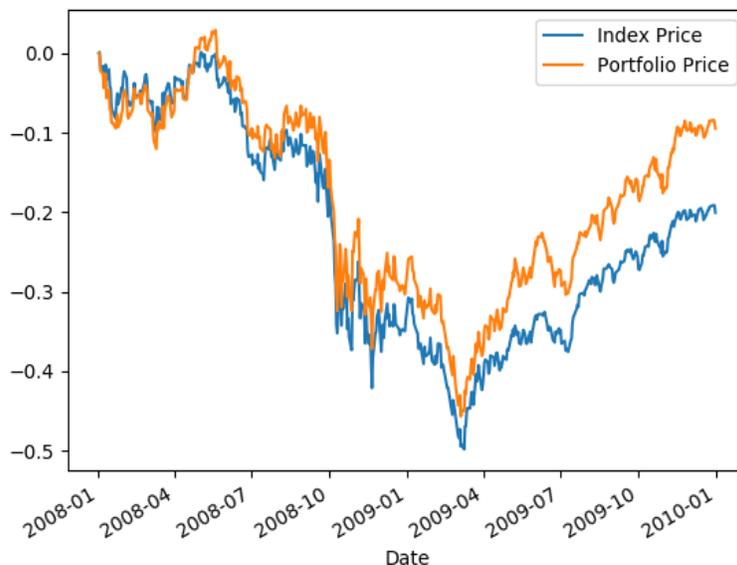

Your undiversified portfolio, which consists of ['XOM', 'AXP', 'MCD', 'BA', 'DIS', 'JNJ', 'HD', 'NKE', 'AAPL', 'DWDP'] , outperforms ^DJI 85.15 % of the time.

Next, we'll move on and attempt to outperform the S&P 500. Although we will not be using all 500 of the index's constituents, we will be using the 50 most weighted ones with at data from at least 2005.

| MSFT | JPM | VZ | WFC | KO | PEP | MDT | MMM | HON | UTX |
|------|------|------|------|------|------|------|------|------|------|
| AAPL | XOM | BAC | INTC | MA | C | DWDP | NFLX | CRM | CVS |
| AMZN | GOOG | PG | CSCO | BA | MCD | AMGN | UNP | MO | TMO |
| BRK-A | UNH | CVX | MRK | CMCSA | WMT | ABT | IBM | ACN | NKE |
| JNJ | PFE | T | HD | DIS | ORCL | ADBE | LLY | COST | TXN |

Inputting the tickers into our program, taking data from January 1st, 2010 to 2014, and finding the ideal threshold (0.475) for a diversified portfolio produced this graph.

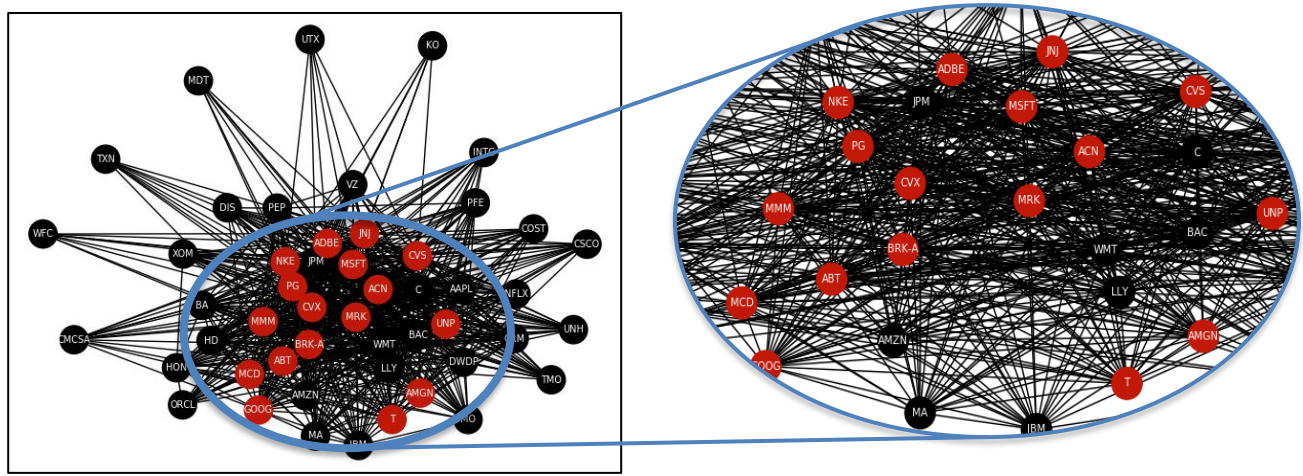

The subgraph that most closely resembles a complete graph consists of 17 out of the original 50 stocks. They are Microsoft (MSFT), Berkshire Hathaway (BRK-A), Johnson & Johnson (JNJ), Alphabet (GOOG), Procter & Gamble (PG), Chevron (CVX), AT&T (T), Merck (MRK), McDonalds (MCD), Amgen (AMGN), Abbot Laboratories (ABT), Adobe (ADBE), 3M (MMM), Union Pacific (UNP), Accenture (ACN), CVS Health (CVS), and Nike (NKE).

From there entering it into the portfolio emulation program, it is visible both from the graph and the percent of times it outperforms the index, that the diversified portfolio does close higher than the index most of the time, but ultimately fails to substantially increase profits.

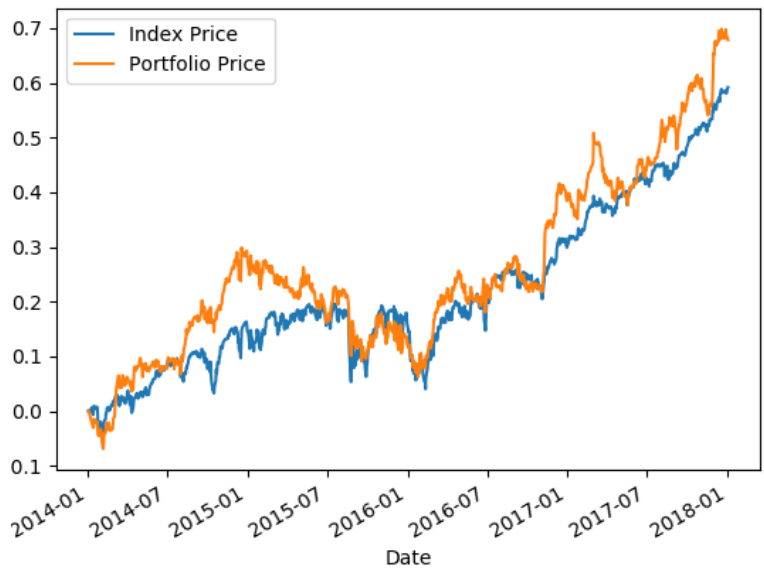

Your diversified portfolio, which consists of ['MSFT', 'BRK-A', 'JNJ', 'GOOG', 'PG', 'CVX', 'T', 'MRK', 'MCD', 'AMGN', 'ABT', 'ADBE', 'MMM', 'UNP', 'ACN', 'CVS', 'NKE'], outperforms SPY 80.85 % of the time.

Different results can be seen with an undiversified portfolio created from the base stocks from the S&P 500, as well as the same timeframe, with an ideal threshold (0.5).

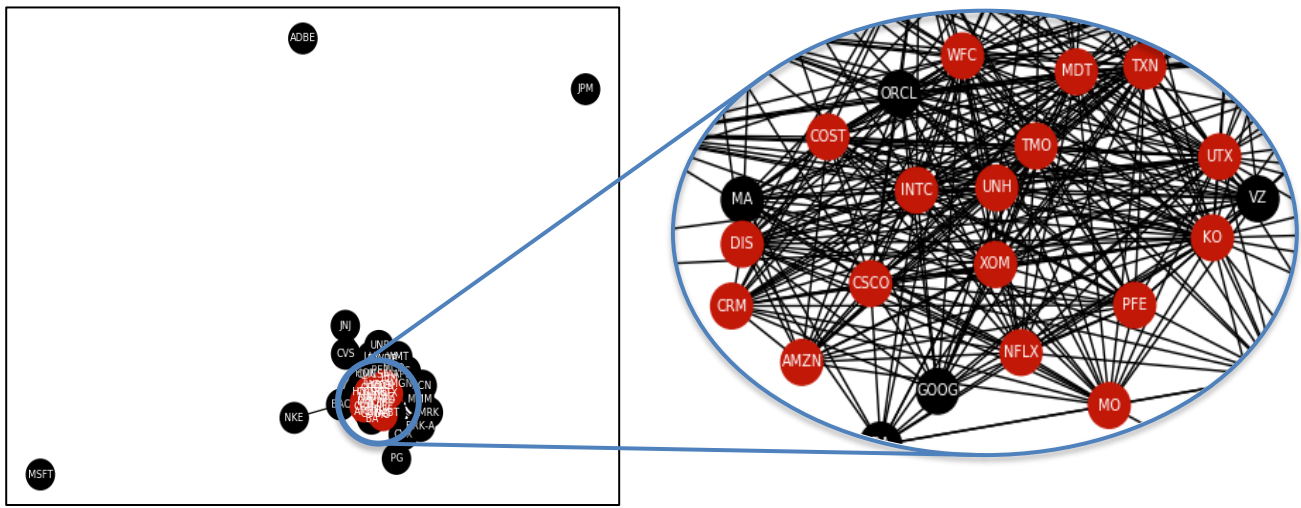

The most optimal subgraph consists of 17 out of the original 50 stocks. The 17 are ExxonMobil (XOM), UnitedHealth (UNH), Pfizer (PFE), Wells Fargo (WFC), Intel (INTC), Cisco Systems (CSCO), Coca-Cola (KO), Disney (DIS), Medtronic (MDT), Netflix (NFLX), salesforce.com (CRM), Altria Group (MO), Costco (COST), United Technologies (UTX), Thermo Fisher (TMO), and Texas Instruments (TXN). Emulating the preceding portfolio yields profits greater than those of the diversified portfolio, but a more volatile investment since it only surpasses the index 69% of the time versus the 81% from the diversified portfolio.

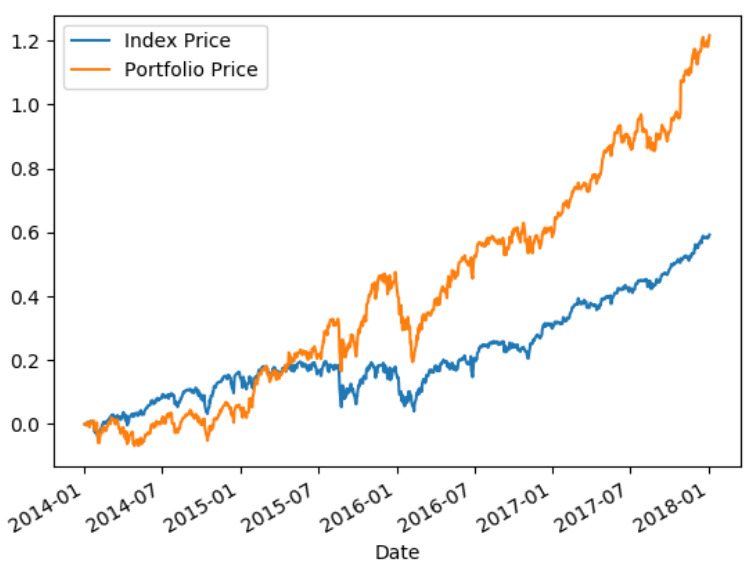

Your undiversified portfolio, which consists of ['AMZN', 'XOM', 'UNH', 'PFE', 'WFC', 'INTC', 'CSCO', 'KO', 'DIS', 'MDT', 'NFLX', 'CRM', 'MO', 'COST', 'UTX', 'TMO', 'TXN'] , outperforms SPY 69.25 % of the time.

Once again, our portfolios fare well in times of economic stability but how would our portfolio creation strategy compare to the S&P 500 during the Great Recession. For this, we would need to use all the original 50 with data ranging back to the beginning of 2005.

An analysis of the least correlated stocks using data from January 1st, 2005 to 2008 and a threshold of 0.323 resulted in this graph.

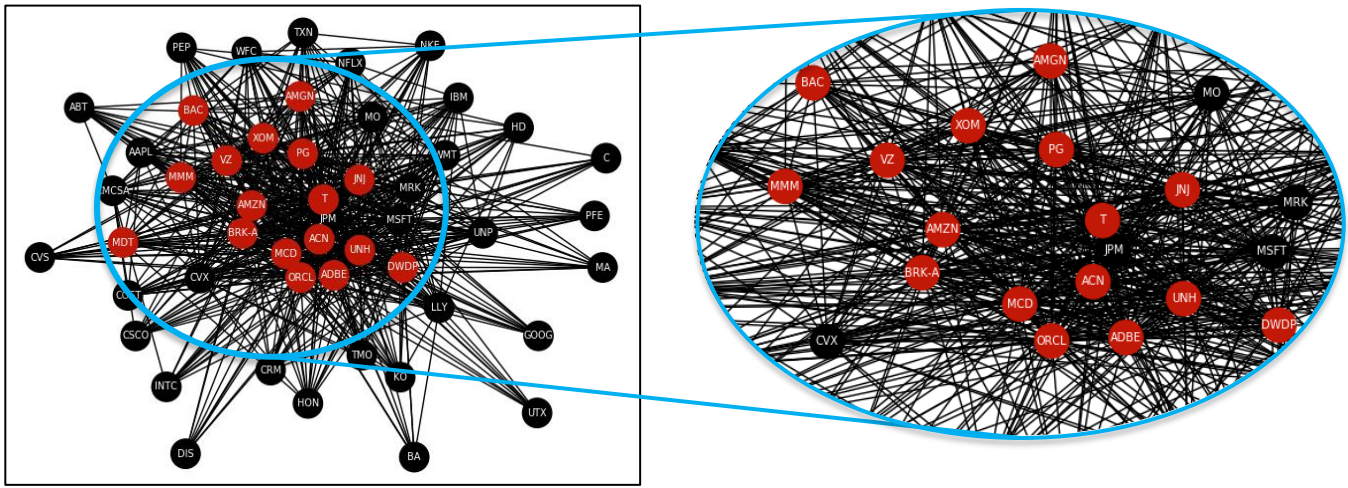

The subgraph that most closely resembles a complete graph consists of 17 out of the original 50 stocks. They are Berkshire Hathaway (BRK-A), Johnson & Johnson (JNJ), ExxonMobil (XOM), UnitedHealth (UNH), Verizon (VZ), Bank of America (BAC), Proctor & Gamble (PG), AT&T (T), McDonalds (MCD), Oracle (ORCL), Medtronic (MDT), DowDuPont (DWDP), Amgen (AMGN), Adobe (ADBE), 3M (MMM), and Accenture (ACN).

The portfolio managed to stay ahead of the index during the Great Recession but before and after struggles to outperform it.

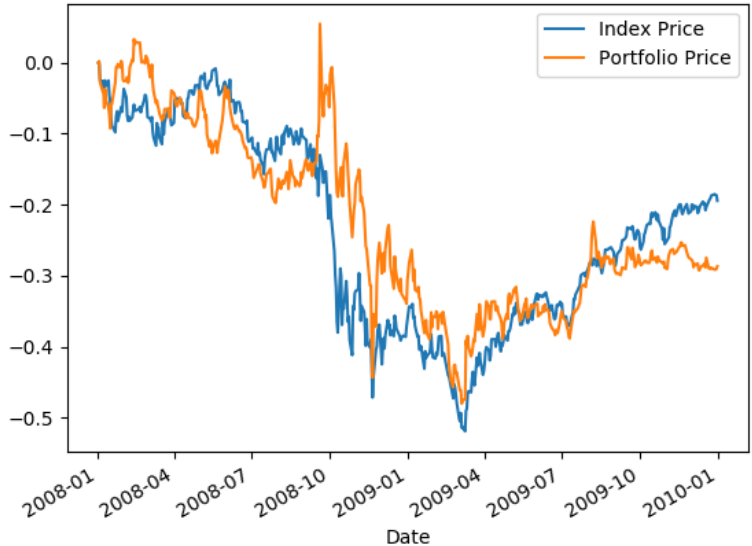

Your diversified portfolio, which consists of ['AMZN', 'BRK-A', 'JNJ', 'XOM', 'UNH', 'VZ', 'BAC', 'PG', 'T', 'MCD', 'ORCL', 'MDT', 'DWDP', 'AMGN', 'ADBE', 'MMM', 'ACN'] , outperforms SPY 48.12 % of the time.

Attempting the same with an undiversified consisting of the most correlated stocks of the original 50 produces a graph when a threshold of 0.34 is used.

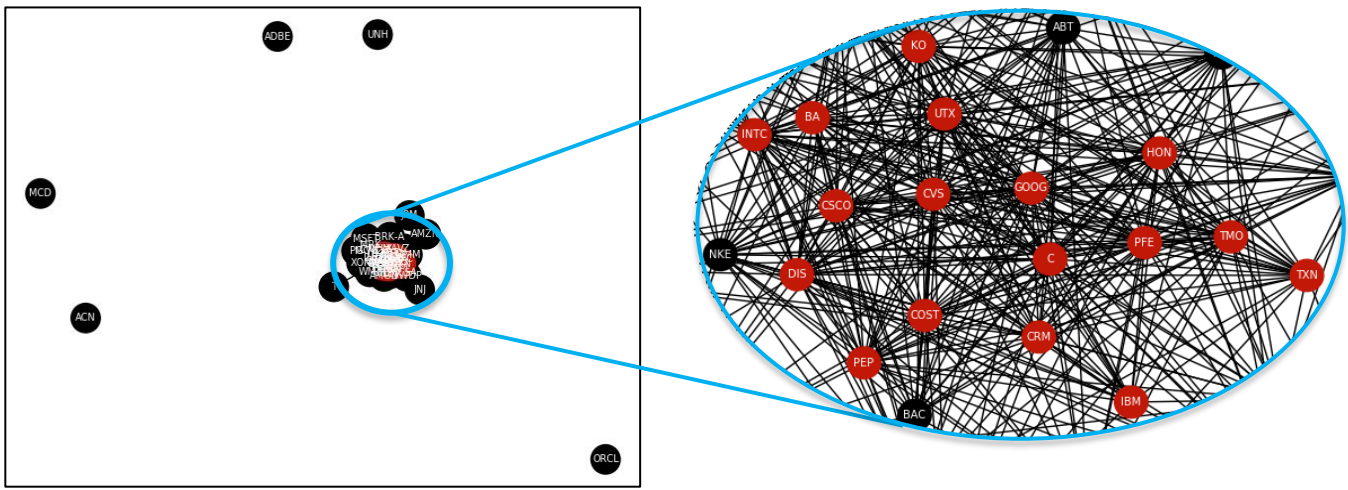

The subgraph that most closely resembles a complete graph consists of 17 out of the original 50 stocks. They are: Alphabet (GOOG), Pfizer (PFE), Intel (INTC), Cisco Systems (CSCO), Coca-Cola (KO), Boeing (BA), Disney (DIS), PepsiCo (PEP), Citigroup (C), International Business Machines (IBM), Honeywell International (HON),

salesforce.com (CRM), Costco (COST), United Technologies (UTX), CVS Health (CVS), Thermo Fisher (TMO), and Texas Instruments (TXN).

The portfolio, unlike the diversified one, falls behind the index during the recession and doesn't recuperate until a couple of years after.

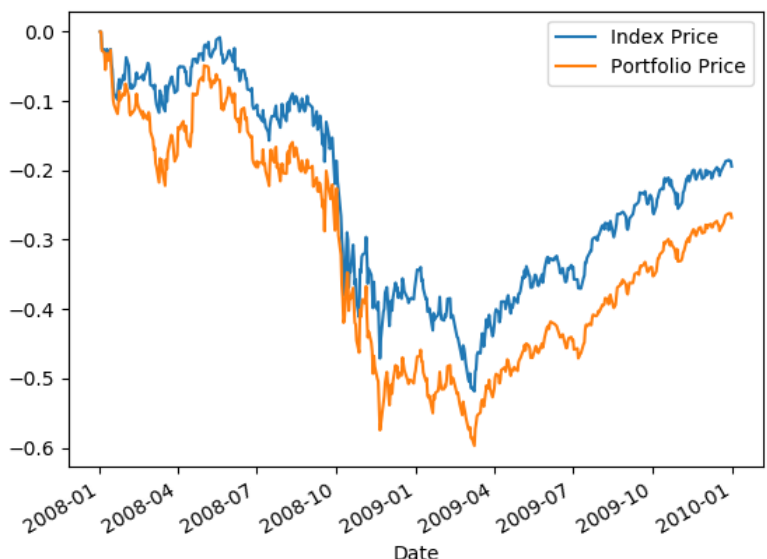

Your portfolio, which consists of ['GOOG', 'PFE', 'INTC', 'CSCO', 'KO', 'BA', 'DIS', 'PEP', 'C', 'IBM', 'HON', 'CRM', 'COST', 'UTX', 'CVS', 'TMO', 'TXN'] , outperforms SPY 0.2 % of the time.

## 5.2. Can we use Stock Y's movement today to predict Stock A's movement n days from today?

### 5.2.1. Program Development

How can we create a program that will determine the best times to invest in a stock? Is it possible to find stocks that have high correlations with the investment stock, and thereafter use it to as an indicator? When the indicator's price moves up, n day's later will the investment stock will also move up? Our first step is to ask the user for the stock that will be invested in and the stocks used to indicate the prime time to invest. We also ask the user for the starting and ending of the data, an ending time for the simulation, and then put it into two separate data tables.

```
18. indicators = []
19.
20. ticker_input = input("Please enter the stock ticker you have chosen to invest in. \n")
21.
22. indicator_input = input("Please enter the stock ticker you have chosen to be in your portfolio.Whe
    n you are done just type 'DONE'! \n")
23.
24. while indicator_input != "DONE":
25.     indicators.append(str(indicator_input))
26.     indicator_input = input()
27.
28. start_date= input("Here enter the date from which data will begin to be taken from (Please put it
    into format of YYYY-MM-DD)\n")
29.
30. data_end_date= input("Here enter the date at which we will stop taking data from (Please put it in
    to format of YYYY-MM-DD)\n")
31.
32. emulation_end_date = input("Here enter the date that the investing simulation will go up to (Pleas
    e put it into format of YYYY-MM-DD)\n")
```

For each of the stocks that will be used as an indicator, we need to find a shift in days that will produce the highest correlation, which will, in turn, produce the most verifiable results. Let's try to understand how this will work. If I want to invest in stock A on January 2 and to indicate when to do so, I use stock B and stock C.

| Date | Stock A (Investment) | Stock B (Indicator) | Stock C (Indicator) |
|---|---|---|---|
| January 1 | 2 | 2 | 9 |
| January 2 | 5 | 3 | 13 |
| January 3 | 11 | 9 | 14 |
| January 4 | 15 | 14 | 16 |
| January 5 | 17 | 15 | 22 |
| January 6 | 24 | 18 | 17 |

With no shift in days, the correlations of each are:

$$Correlation(A, B) = 0.9707$$
$$Correlation(A, C) = 0.6081$$

Now let's see the correlation of Stock A today (January 2) with Stock B and C yesterday (January 1).

$$Correlation(A, B) = 0.922$$
$$Correlation(A, C) = 0.9845$$

So, Stock A and B have the highest correlation without a shift in days. Stock C today, however, has a very high correlation with stock A the day after. Let's use our correlation function from the previous program and incorporate a shift in days. On top of that lets add a dictionary that will record both the number of days that data must be shifted to produce the highest correlation and the correlation itself.

```
39. def corr(x,y,days):
40.     x_avg = x.mean()
41.     y_avg = y.mean()
42.     numerator_sum = 0
43.     denominator_sum_x = 0
44.     denominator_sum_y = 0
45.     for i in range(days,len(x)):
46.         numerator_sum += (x[i] - x_avg)*(y[i-days] - y_avg)
47.         denominator_sum_x += (x[i] - x_avg) ** 2
48.         denominator_sum_y += (y[i-days] - y_avg) ** 2
49.     correlation_value = (numerator_sum / ((denominator_sum_x*denominator_sum_y) ** (0.5)))
50.     return correlation_value
51.
52. OPTIMAL_SHIFT_DICTIONARY = {}
53. OPTIMAL_CORR_DICTIONARY = {}
54.
55. ticker_emulation_data = pdr.DataReader(ticker_input, 'yahoo', data_end_date, emulation_end_date)['Adj Close']
56. indicator_emulation_data = pdr.DataReader(indicators, 'yahoo', data_end_date, emulation_end_date)['Adj Close']
57.
58. for indicator_index in range (0,len(indicators)):
59.     highest_correlation = corr(ticker_data, indicator_data[indicators[indicator_index]],1)
60.     highest_days = 1
61.     for n in range (1,80):
62.         if corr(ticker_data, indicator_data[indicators[indicator_index]],n) > highest_correlation:
63.             highest_correlation = corr(ticker_data, indicator_data[indicators[indicator_index]],n)
64.             if n > highest_days:
65.                 highest_days = n
66.
67.     OPTIMAL_SHIFT_DICTIONARY[indicators[indicator_index]] = highest_days
68.     OPTIMAL_CORR_DICTIONARY[indicators[indicator_index]] = highest_correlation
```

Next, we will have to create a program that utilizes the prime relationships recently recorded, and for each day check the validity of those conditions. If the conditions are true, the stock will be invested in. The user is also given an option of how many of the conditions must be true in order to invest.

```
76.  def test(r):
77.      conditionals = 0
78.      for indicator_index in range (0,len(indicators)):
79.          if indicator_emulation_data_pct[indicators[indicator_index]][r-
    OPTIMAL_SHIFT_DICTIONARY[indicators[indicator_index]]] > 0:
80.              conditionals += 1
81.      if conditionals >= number_of_true:
82.          return "true"
83.      else:
84.          return "false"
85.
86.
87.  portfolio_price = ticker_emulation_data[0]
88.
89.  print(len(indicators),"relationships have been developed. How many of these relationships must be true in order to invest in"
90.        ,ticker_input,"To try a new number close window and type in a new number. If you are done en
    ter any negative number.")
91.  number_of_true = int(input())
92.
93.  while number_of_true >= 0:
94.      continuous_investing = ticker_emulation_data[:]
95.      indicative_investing = []
96.      for r in range (0,len(ticker_emulation_data.index)):
97.          if number_of_true == 0:
98.              if r == 0:
99.                  last_true = portfolio_price
100.             if r != 0:
101.                 last_true = (last_true+ (ticker_emulation_data[r]-ticker_emulation_data[r-1]))
102.         else:
103.             if r < highest_days:
104.                 last_true = portfolio_price
105.             if r >= highest_days:
106.                 if test(r) == "true":
107.                     last_true = (last_true+ (ticker_emulation_data[r]-ticker_emulation_data[r-
    1]))
108.         indicative_investing.append(last_true)
```

Finally, in the end, a directed graph will display that will show the user the conditional results for the last day in the simulation date. This will help the user decide if they will invest the following day. Below is the program that creates and displays the directed graph for the last day.

```
120. labels={}
121. labels[0] = ticker_input
122. for i in range (1,len(indicators)):
123.     labels[i] = indicators[i]
124.
125. graph = {}
126.
127. for indicator_index in range (0,len(indicators)):
128.     if indicator_emulation_data_pct[indicators[indicator_index]][(len(ticker_emulation_data.index)
    )-OPTIMAL_SHIFT_DICTIONARY[indicators[indicator_index]]] > 0:
129.         graph[indicators[indicator_index]] = '1'
```

```
130.
131.
132. labels = {}
133. labels['1'] = ticker_input
134.
135. print(graph)
136. G = nx.DiGraph(graph)
137. G.add_nodes_from(indicators)
138. H = nx.relabel_nodes(G,labels)
139. nx.draw(H,node_color='black',with_labels=True,node_size=800,font_size=10,font_color="white")
140. plt.show()
```

This graph is what the program does for every day of data and the positive movement of indicator stocks represent directed edges. We will refer to the positive movement of indicator stocks n days prior as a conditional. If the indicator stock had a positive movement the conditional is true and therefore an edge is present. That also means that the number of true conditionals is the measure of the degree of the center node/investment stock.

### 5.2.2 Testing Effectiveness

To test the effectiveness of investing based off indicative stocks, we will need to backtest the method. For our first test, we will be investing in MSFT, the largest component of the SPY, and as indicators we will use the next 4 largest components of the SPY, each from a different sector and with at least 10 years of data, Amazon (AMZN), Berkshire Hathaway (BRK-A), Johnson & Johnson (JNJ), and Alphabet (GOOG).

For our first trial, we will take data from January 1$^{st,}$ 2010 to 2017 to find relationships and emulate investing from January 1$^{st}$, 2017 until December 27$^{th}$, 2017. Data will be shifted up to 79 days in order to find the highest correlation, the best relationship.

The best relationships in our range of data shifts are:
Today's MSFT price has the highest correlation (0.93) with AMZN's price 1 day before

Today's MSFT price has the highest correlation (0.92) with BRK-A's price 1 day before

Today's MSFT price has the highest correlation (0.95) with JNJ's price 12 days before

Today's MSFT price has the highest correlation (0.96) with GOOG's price 1 day before

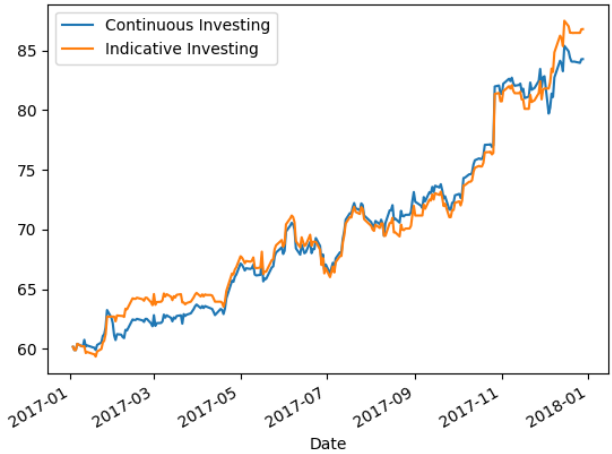
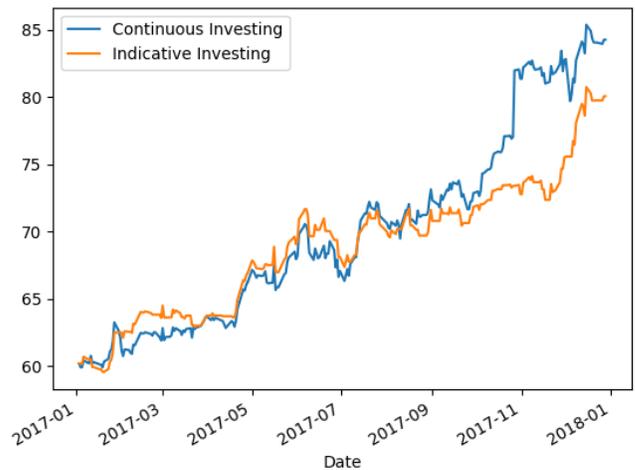
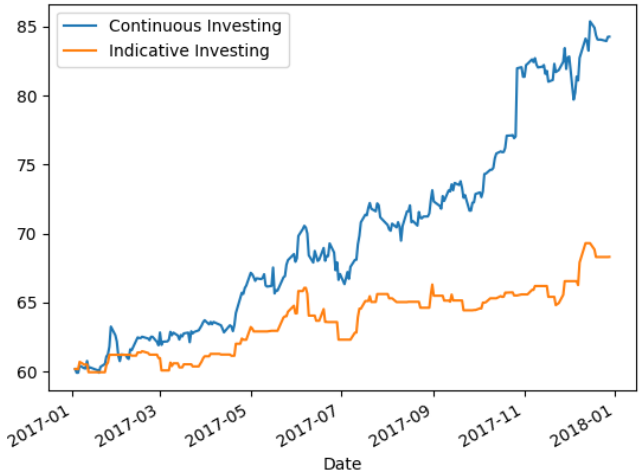
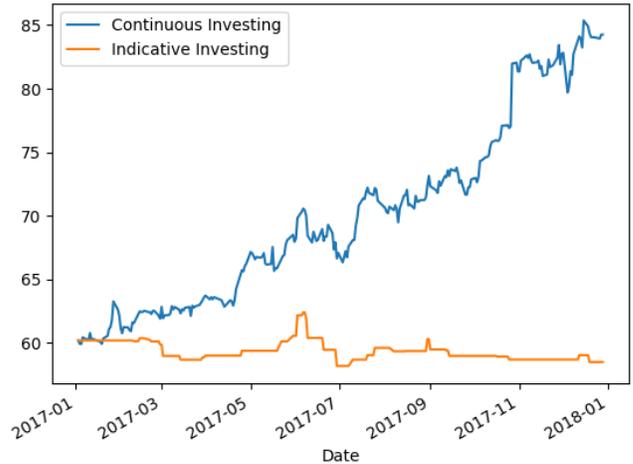

For investors, the following graph represents how the conditions are met to indicate whether they should invest the following day, December 28th, 2017.

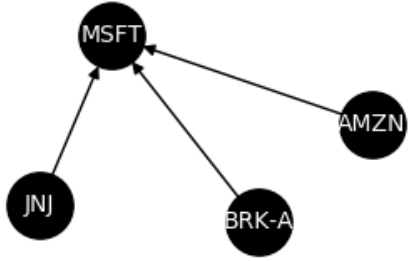

As we can see 3 conditions are met so it's likely that MSFT would increase in price on the 28th. Nonetheless, using indicative investing doesn't outperform at all. Maybe the fact that we used so much data to establish the correlations means that the correlations aren't as precise since correlation does change over time. What if we start taking data at the start of 2015, instead of 2010?

The best relationships in our range of data shifts are:

Today's MSFT price has the highest correlation (0.89) with AMZN's price 14 days before

Today's MSFT price has the highest correlation (0.33) with BRK-A's price 1 day before

Today's MSFT price has the highest correlation (0.78) with JNJ's price 1 day before

Today's MSFT price has the highest correlation (0.90) with GOOG's price 1 day before

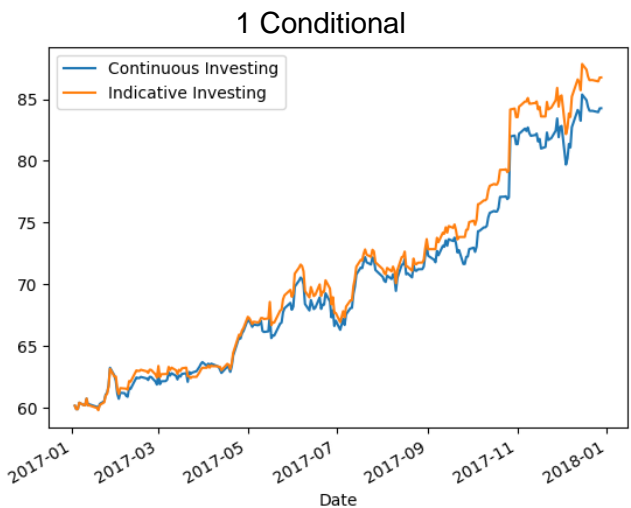
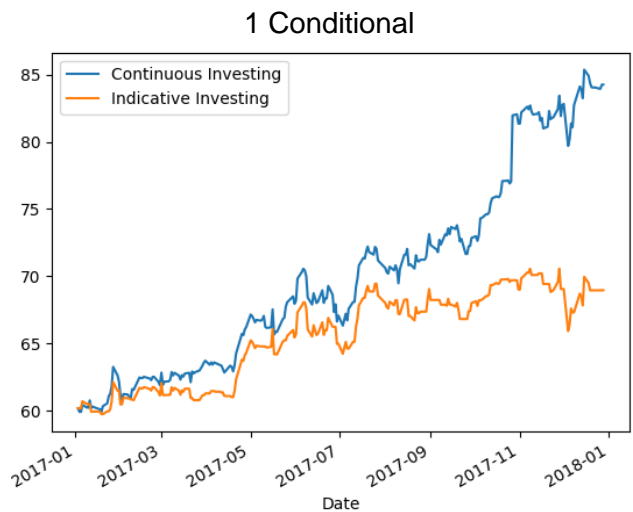

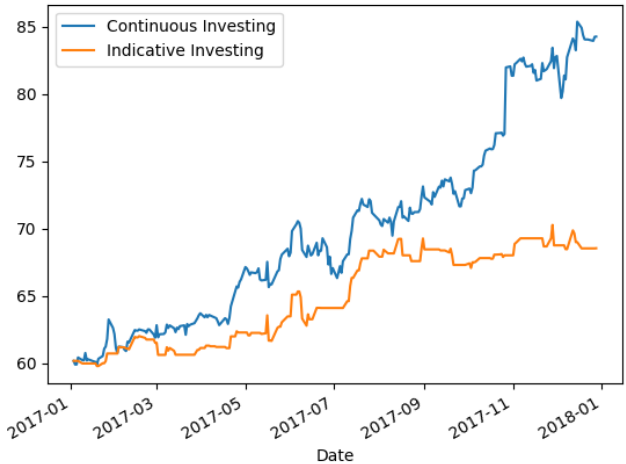
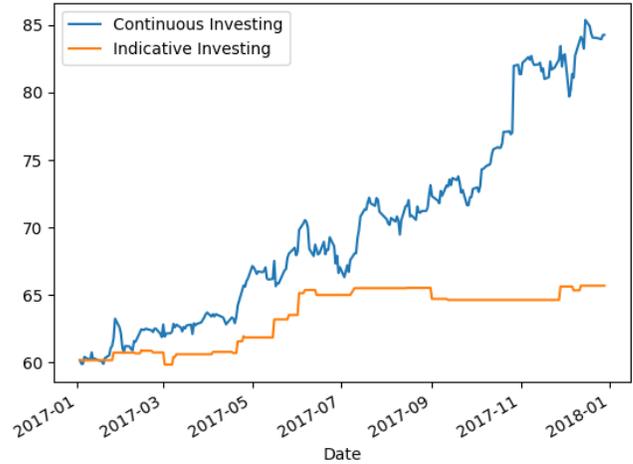

For investors, the following graph represents how the conditions are met to indicate whether they should invest the following day, December 28th, 2017.

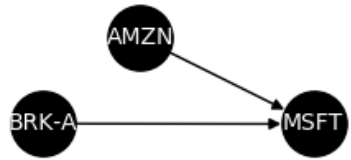

Only two conditions are met, and the correlation between MSFT and BRK-A is only 0.33, so an investor most likely would not invest in MSFT on January 28th from this data. With this trial, investing when one conditional was true did outperform continuous investing most of the time, and two conditionals outperformed continuous investing for the first half of the year then fell behind.

Let's repeat this with 5 random stocks. We'll invest in Capital City Bank Group (CCBG), and use Summit State Bank (SSBI), Immunomedics (IMMY), Craft Brew Alliance (BREW), and Pebble brook Hotels (PEB) as indicators. Again, we will take data from January 1st, 2010 until January 1st, 2017 to find relationships and emulate investing from January 1st, 2017 until December 27th, 2017.

The best relationships in our range of data shifts are:

Today's CCBG price has the highest correlation (0.805) with SSBI's price 79 days before

Today's CCBG price has the highest correlation (0.02) with IMMY's price 52 days before

Today's CCBG price has the highest correlation (0.52) with BREW's price 74 days before

Today's CCBG price has the highest correlation (0.69) with PEB's price 79 days before

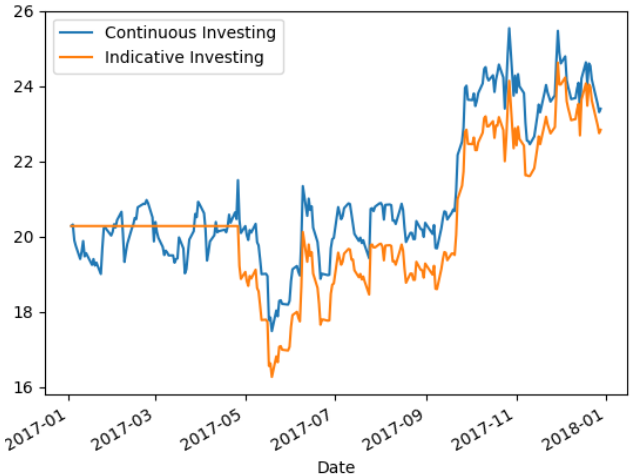
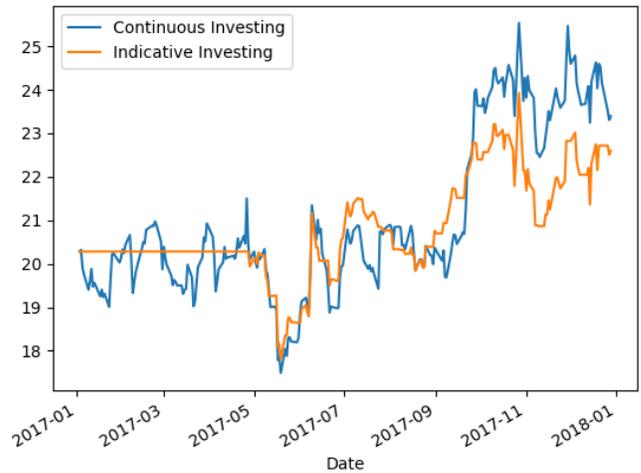
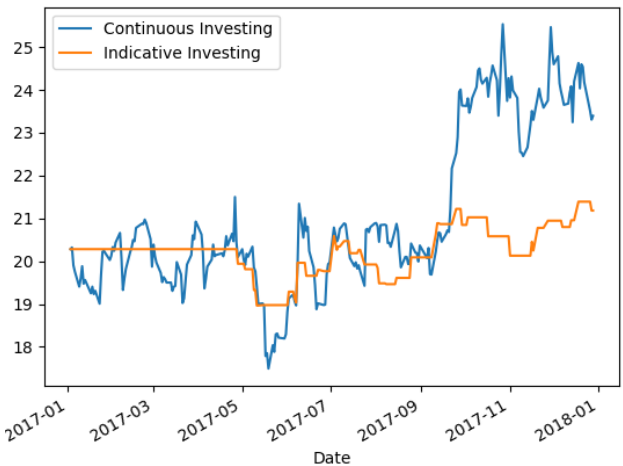
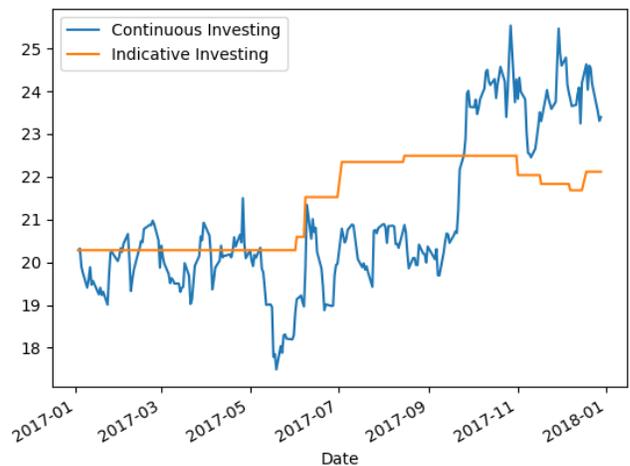

For investors, the following graph represents how the conditions are met to indicate whether they should invest the following day, December 28th, 2017.

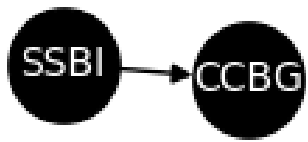

As we can see, no conditions are met an investor should be very hesitant to invest the following day. With all the indicative investing except for the two-conditional investing, perform much worse than continuously investing. When two conditions are true, investment sometimes outperforms continuous investing for the first half of the year. Let's try it again but with data from 2015.

The best relationships in our range of data shifts are:

Today's CCBG price has the highest correlation (0.479) with SSBI's price 1 day before

Today's CCBG price has the highest correlation (-0.22) with IMMY's price 63 days before

Today's CCBG price has the highest correlation (0.71) with BREW's price 64 days before

Today's CCBG price has the highest correlation (0.017) with PEB's price 1 day before

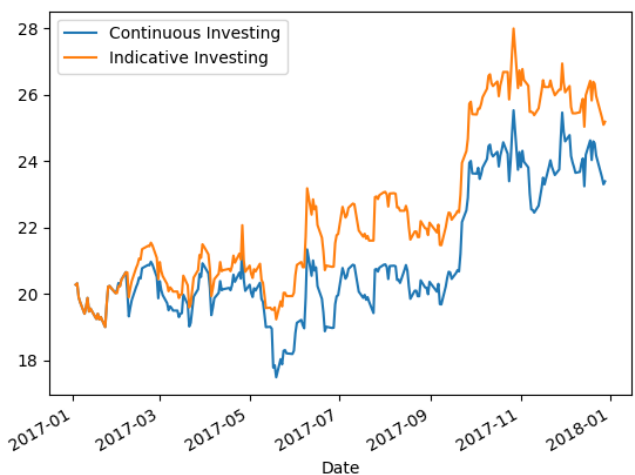
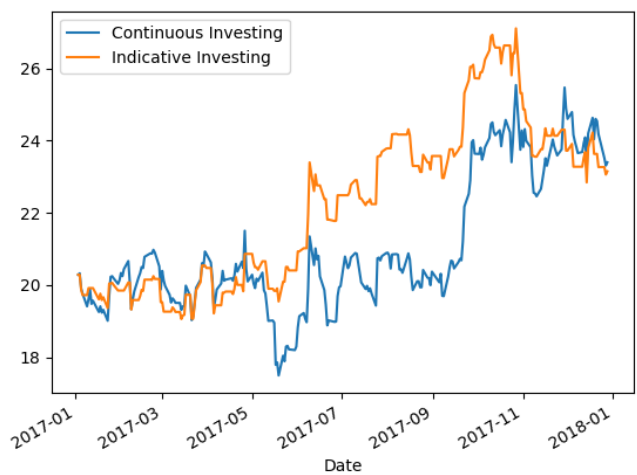

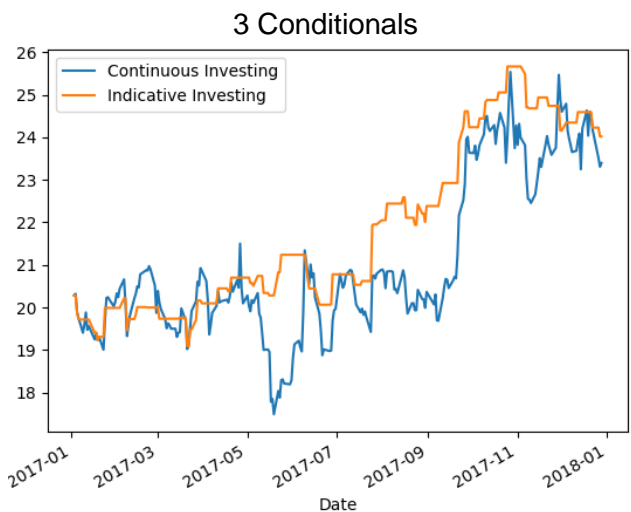 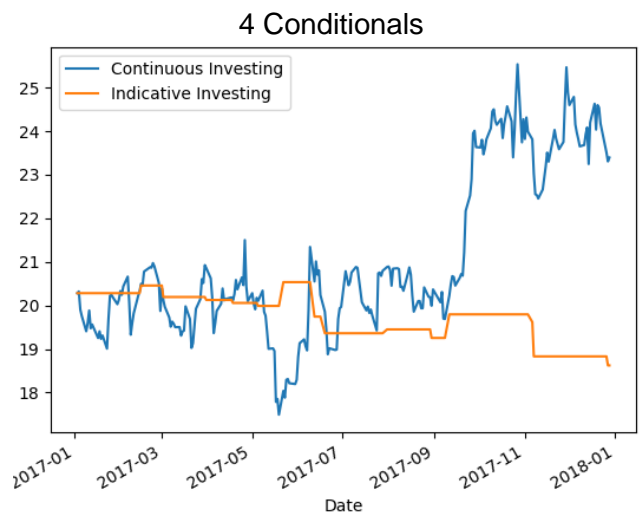

For investors, the following graph represents how the conditions are met to indicate whether they should invest the following day, December 28th, 2017.

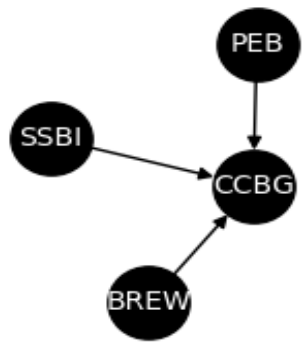

Interesting! Even though the correlations were lower, or even negative, the indicative investing did better than continuous investing for both one and two conditionals being true. When one was true, the investment always outperformed continuous investing, but for two it only flew ahead at a few points.

# 6. Conclusions

The applications of graph theory to investing are extensive but the in-depth exploration of two methods showed varied results.

Initially, we explored transforming correlation matrices into adjacency matrices in order to find a portfolio of stocks in which every pair has either low correlations (diversified) or high correlations (undiversified). The diversified portfolio consistently outperforms the index during economic stability, although during the Great Recession in 2008 is unsuccessful in minimizing losses. The undiversified portfolio yields erratic results in which the portfolio either races ahead or falls behind the index.

Then we investigated how directed graphs can help us understand when the best time to invest in a stock is using optimized relationships based on data shifting. Once we know that stock A has the highest correlation with stock B 3 days prior and stock C 17 days prior, for every day we can create a directed graph in which a positive increase in stock B and stock C those amount of days before has positive returns thus indicating that it is likely that stock A will also have positive returns. Unfortunately, this method shows little success when basing indicator stocks off having an extremely high correlation but showed promising success (around 10% increase in profits) when utilizing a random set that had various types of correlations (negative, weak positive, strong positive) with the investment stock. The success of this method also seems to improve when the data is taken from a smaller set due to a more precise and relevant correlation.

# 7. Applications and Extensions

The principles discussed in this paper are extremely applicable to a huge audience. The contents of this paper and utilized skills can be used in order to increase financial literacy amongst students. There is a huge lack of financial literacy and part of the purpose of writing this paper was to allow peers to understand the basics of investing and portfolio development. The process of developing portfolios and programs used to do so can help everyday investors as well as professional traders. For example, if someone wanted to start a retirement fund in 2014 for when they retire in 2018, they would typically choose to put it into a 401k plan which is invested into the S&P 500. If they followed the methods discussed in **5.1.2.**, they would have an extra 10% return from their investment in our portfolio rather than the S&P. The results are even more drastic if instead of investing in the Dow Jones Index in anticipation of retirement, they invested in the portfolio based off it which results in over 30% more profit over the same 4-year period.

This research can be further extended by attempting to increase success rates of trials. this can be done by incorporating moving correlations into the programs and data analysis. Like how a smaller dataset resulted in a more precise correlation in **5.2.2.**, moving correlations will account for fundamental changes in stock relationships and thus will be able to say that during 2017 stock A and B had a correlation of -0.2, but during 2018 had a correlation of 0.63. This would be more helpful than giving a value between -0.2 and 0.63 for 2017-2018 that wouldn't accurately describe the relationship between the stocks for either year. Another possible extension would be to dive deeper into the exploration of using indicators to determine when to invest in a stock. What would make a good indicator, and what made the 4 random stocks chosen in our last trial in **5.2.2.** effective?

# 9. Appendices

## A. Correlation Graph Program

```python
"""
Install:
$ pip install fix_yahoo_finance --upgrade --no-cache-dir
pip install pandas-datareader
pip install networkx
"""

from pandas_datareader import data as pdr
import pandas as pd
import numpy as np
import fix_yahoo_finance as yf
import matplotlib.pyplot as plt
import statistics as statistics
import networkx as nx

yf.pdr_override()

tickers = []

ticker_input = input("Please enter the stock tickers you would like to use one by one. When you are done just type 'DONE'! \n")

while ticker_input != "DONE":
    tickers.append(str(ticker_input))
    ticker_input = input()

start_date = input("Here enter the date from which data will begin to be taken from (Please put it into the format of YYYY-MM-DD)\n")

end_date = input("Here enter the date at which we will stop taking data from (Please put it into the format of YYYY-MM-DD)\n")

#Downloads data from Yahoo
data = pdr.DataReader(tickers, 'yahoo', start_date, end_date)['Adj Close']

#Changes values to percent change for uniform measurements
data = data.pct_change()

#Removes first percent change and missing data as its NaN
data = data.dropna(axis=0, how='any', thresh=None, subset=None, inplace=False)

data = data.values

x_avg = 0
y_avg = 0

#Defines function correlation that can be used to find correlation between two different data sets for different stocks
def corr(x,y):
    x_avg = x.mean()
```

```python
48.        y_avg = y.mean()
49.        numerator_sum = 0
50.        denominator_sum_x = 0
51.        denominator_sum_y = 0
52.        for i in range(len(x)):
53.            numerator_sum += (x[i] - x_avg)*(y[i] - y_avg)
54.            denominator_sum_x += (x[i] - x_avg) ** 2
55.            denominator_sum_y += (y[i] - y_avg) ** 2
56.        correlation_value = (numerator_sum / ((denominator_sum_x*denominator_sum_y) ** (0.5)))
57.        return correlation_value
58.
59.
60. correlations = np.zeros((len(tickers),len(tickers)))
61. correlations_list = []
62.
63. #Makes correlation matrix for all inputted stock tickers at beginning of program
64. for ticker in range(0,len(tickers)):
65.     for tickerpair in range(0,len(tickers)):
66.         if ticker != tickerpair:
67.             correlations[ticker,tickerpair] = corr(data[:,ticker],data[:,tickerpair])
68.             correlations_list.append(corr(data[:,ticker],data[:,tickerpair]))
69.
70. print("\n",correlations)
71.
72. #Prints the average correlation for the user
73. print("\nThe average correlation is",statistics.mean(correlations_list))
74.
75. #Prints the average correlation for the user
76. print("\nThe median correlation is",statistics.median(correlations_list))
77.
78. adjacency_matrix = np.zeros((len(tickers),len(tickers)))
79.
80. standard_deviation_sum = 0
81.
82. #Finds and utilizes standard deviation to suggest a threshold to the user for the given dataset
83. for i in range(0,len(correlations_list)):
84.     standard_deviation_sum += (correlations_list[i] - statistics.mean(correlations_list)) ** 2
85. standard_deviation = ((standard_deviation_sum / (len(correlations_list) - 1)) ** (0.5))
86.
87. #Asks the user if they are looking for correlated stocks or uncorrelated stocks.
88. user_preference = input("Type in 'U' if you want undiversified(correlated) stocks in your portfolio, or 'D' if you want diversified(uncorrelated) stocks.")
89.
90.
91. if user_preference == "D":
92.     #Subtracts one standard deviation from the mean to get the 16 percent most uncorrelated relationships
93.     threshold_with_standard_deviation = (statistics.mean(correlations_list)
94.                                         - standard_deviation)
95.
96. if user_preference == "U":
97.     #Adds one standard deviation from the mean to get the 16 percent most correlated relationships
98.     threshold_with_standard_deviation = (statistics.mean(correlations_list)
99.                                         + standard_deviation)
100.
101. print("\nUsing the Standard Deviation,", threshold_with_standard_deviation," should be the best threshold for the inputted stock tickers.")
102.
103. threshold = float(input("\nInput a threshold. To change threshold just close the graph and type in a new threshold.\n"))
```

```
104.    #Makes the previous correlation matrix an adjacency matrix for stocks an takes into account the co
        rrelation and threshold.
105.    while threshold != "DONE":
106.        for ticker in range(0,len(tickers)):
107.            for tickerpair in range(0,len(tickers)):
108.                if ticker != tickerpair:
109.                    if user_preference == "D":
110.                        if abs(corr(data[:,ticker],data[:,tickerpair])) < threshold:
111.                            adjacency_matrix[ticker,tickerpair] = 1
112.                        if abs(corr(data[:,ticker],data[:,tickerpair])) > threshold:
113.                            adjacency_matrix[ticker,tickerpair] = 0
114.                    if user_preference == "U":
115.                        if abs(corr(data[:,ticker],data[:,tickerpair])) > threshold:
116.                            adjacency_matrix[ticker,tickerpair] = 1
117.                        if abs(corr(data[:,ticker],data[:,tickerpair])) < threshold:
118.                            adjeaency_matrix[ticker,tickerpair] = 0
119.
120.        print("\n",adjecency_matrix)
121.
122.        labels={}
123.        for i in range (0,len(tickers)):
124.            labels[i] = tickers[i]
125.
126.        # Draws and displays the graph
127.        G=nx.Graph(adjacency_matrix)
128.        H=nx.relabel_nodes(G,labels)
129.        complete_graphs = [s for s in nx.enumerate_all_cliques(H) if len(s) > 1]
130.        max_complete_graph = complete_graphs[len(complete_graphs)-1]
131.        print(max_complete_graph)
132.
133.        color_map = []
134.        for index in range (0,len(tickers)):
135.            if tickers[index] in max_complete_graph:
136.                color_map.append('#C21807')
137.            if tickers[index] not in max_complete_graph:
138.                color_map.append('black')
139.
140.        nx.draw(H,node_color=color_map,with_labels=True,node_size=450,font_size=7,font_color="white")
141.        plt.axis('off')
142.        plt.show()
143.        threshold= float(input())
```

### B.    Portfolio Emulation Program

```
1.  """
2.  Simulates Portfolio to gauge effectiveness
3.
4.  Install:
5.  $ pip install fix_yahoo_finance --upgrade --no-cache-dir
6.  pip install pandas-datareader
7.  """
8.  from pandas_datareader import data as pdr
9.  import matplotlib.pyplot as plt
10. import pandas as pd
11. import numpy as np
12. import fix_yahoo_finance as yf
13.
14.
15. yf.pdr_override()
```

```
16.  
17.  tickers = []
18.  
19.  ticker_input = input("Please enter the stock tickers you have chosen to be in your portfolio.When
     you are done just type 'DONE'! \n")
20.  
21.  while ticker_input != "DONE":
22.      tickers.append(str(ticker_input))
23.      ticker_input = input()
24.  
25.  start_date= input("Here enter the date from which data will begin to be taken from (Please put it
     into format of YYYY-MM-DD)\n")
26.  
27.  end_date= input("Here enter the date at which we will stop taking data from (Please put it into fo
     rmat of YYYY-MM-DD)\n")
28.  
29.  #Downloads data from Yahoo using inputs given by the user.
30.  data = pdr.DataReader(tickers, 'yahoo', start_date, end_date)['Adj Close']
31.  
32.  print(data)
33.  
34.  WEIGHT_DICTIONARY = {}
35.  WEIGHT_DICTIONARY["MSFT"] = 85492740
36.  WEIGHT_DICTIONARY["AAPL"] = 51156136
37.  WEIGHT_DICTIONARY["AMZN"] = 4567667
38.  WEIGHT_DICTIONARY["BRK-A"] = 21734176
39.  WEIGHT_DICTIONARY["JNJ"] = 29909924
40.  WEIGHT_DICTIONARY["JPM"] = 37470536
41.  WEIGHT_DICTIONARY["CELG"] = 7844412
42.  WEIGHT_DICTIONARY["XOM"] = 47202584
43.  WEIGHT_DICTIONARY["GOOG"] = 3432555
44.  WEIGHT_DICTIONARY["UNH"] = 10730623
45.  WEIGHT_DICTIONARY["PFE"] = 65356490
46.  WEIGHT_DICTIONARY["D"] = 7291978
47.  WEIGHT_DICTIONARY["VZ"] = 46066884
48.  WEIGHT_DICTIONARY["BAC"] = 103563400
49.  WEIGHT_DICTIONARY["PG"] = 27751676
50.  WEIGHT_DICTIONARY["CVX"] = 21363044
51.  WEIGHT_DICTIONARY["T"] = 80963704
52.  WEIGHT_DICTIONARY["WFC"] = 48325410
53.  WEIGHT_DICTIONARY["INTC"] = 51407956
54.  WEIGHT_DICTIONARY["CSCO"] = 50965776
55.  WEIGHT_DICTIONARY["MRK"] = 29650844
56.  WEIGHT_DICTIONARY["HD"] = 12755844
57.  WEIGHT_DICTIONARY["KO"] = 42673950
58.  WEIGHT_DICTIONARY["MA"] = 10171179
59.  WEIGHT_DICTIONARY["BA"] = 5956660
60.  WEIGHT_DICTIONARY["CMCSA"] = 50940380
61.  WEIGHT_DICTIONARY["DIS"] = 16581456
62.  WEIGHT_DICTIONARY["PEP"] = 15768362
63.  WEIGHT_DICTIONARY["C"] = 28057804
64.  WEIGHT_DICTIONARY["MCD"] = 8649315
65.  WEIGHT_DICTIONARY["WMT"] = 15999681
66.  WEIGHT_DICTIONARY["ABBV"] = 16882740
67.  WEIGHT_DICTIONARY["PM"] = 17331124
68.  WEIGHT_DICTIONARY["ORCL"] = 31510256
69.  WEIGHT_DICTIONARY["MDT"] = 15056564
70.  WEIGHT_DICTIONARY["DWDP"] = 25724776
71.  WEIGHT_DICTIONARY["AMGN"] = 7216490
72.  WEIGHT_DICTIONARY["ABT"] = 19558848
73.  WEIGHT_DICTIONARY["ADBE"] = 5459268
```

```
74.  WEIGHT_DICTIONARY["MMM"] = 6533138
75.  WEIGHT_DICTIONARY["NFLX"] = 4854932
76.  WEIGHT_DICTIONARY["UNP"] = 8244006
77.  WEIGHT_DICTIONARY["IBM"] = 10176588
78.  WEIGHT_DICTIONARY["LLY"] = 10645048
79.  WEIGHT_DICTIONARY["HON"] = 8279166
80.  WEIGHT_DICTIONARY["CRM"] = 8436445
81.  WEIGHT_DICTIONARY["MO"] = 20993784
82.  WEIGHT_DICTIONARY["ACN"] = 7143811
83.  WEIGHT_DICTIONARY["AVGO"] = 4812683
84.  WEIGHT_DICTIONARY["COST"] = 4884066
85.  WEIGHT_DICTIONARY["PYPL"] = 13196811
86.  WEIGHT_DICTIONARY["UTX"] = 9043813
87.  WEIGHT_DICTIONARY["CVS"] = 14476559
88.  WEIGHT_DICTIONARY["TMO"] = 4485843
89.  WEIGHT_DICTIONARY["NKE"] = 14260508
90.  WEIGHT_DICTIONARY["TXN"] = 10839165
91.  WEIGHT_DICTIONARY["NVDA"] = 6778510
92.  WEIGHT_DICTIONARY["GILD"] = 14452690
93.  WEIGHT_DICTIONARY["BKNG"] = 529460
94.  WEIGHT_DICTIONARY["BMY"] = 18200668
95.  WEIGHT_DICTIONARY["NEE"] = 5259724
96.  WEIGHT_DICTIONARY["SBUX"] = 13814848
97.  WEIGHT_DICTIONARY["USB"] = 17076010
98.  WEIGHT_DICTIONARY["COP"] = 12956442
99.  WEIGHT_DICTIONARY["AXP"] = 7874480
100. WEIGHT_DICTIONARY["AMT"] = 4914949
101. WEIGHT_DICTIONARY["CAT"] = 6618982
102. WEIGHT_DICTIONARY["UPS"] = 7733609
103. WEIGHT_DICTIONARY["ANTM"] = 2899479
104. WEIGHT_DICTIONARY["LOW"] = 9044233
105. WEIGHT_DICTIONARY["LMT"] = 2762399
106. WEIGHT_DICTIONARY["WBA"] = 9386737
107. WEIGHT_DICTIONARY["QCOM"] = 13514317
108. WEIGHT_DICTIONARY["CME"] = 3940956
109. WEIGHT_DICTIONARY["MDLZ"] = 16363128
110. WEIGHT_DICTIONARY["DUK"] = 7946115
111. WEIGHT_DICTIONARY["BIIB"] = 2245851
112. WEIGHT_DICTIONARY["GS"] = 3916273
113. WEIGHT_DICTIONARY["BDX"] = 2984299
114. WEIGHT_DICTIONARY["DHR"] = 6867878
115. WEIGHT_DICTIONARY["ADP"] = 4886452
116. WEIGHT_DICTIONARY["GE"] = 96935416
117. WEIGHT_DICTIONARY["CB"] = 5166407
118. WEIGHT_DICTIONARY["EOG"] = 6460373
119. WEIGHT_DICTIONARY["SLB"] = 15436883
120. WEIGHT_DICTIONARY["PNC"] = 5178182
121. WEIGHT_DICTIONARY["SPG"] = 3448428
122. WEIGHT_DICTIONARY["TJX"] = 13968048
123. WEIGHT_DICTIONARY["CHTR"] = 1990736
124. WEIGHT_DICTIONARY["ISRG"] = 1268706
125. WEIGHT_DICTIONARY["CSX"] = 9093531
126. WEIGHT_DICTIONARY["MS"] = 14786210
127. WEIGHT_DICTIONARY["CL"] = 9663777
128. WEIGHT_DICTIONARY["ESRX"] = 6268218
129. WEIGHT_DICTIONARY["INTU"] = 2886636
130. WEIGHT_DICTIONARY["SYK"] = 3461191
131. WEIGHT_DICTIONARY["FOXA"] = 11751387
132. WEIGHT_DICTIONARY["OXY"] = 8525703
133. WEIGHT_DICTIONARY["CI"] = 2710333
134. WEIGHT_DICTIONARY["SCHW"] = 13411722
```

```python
135.
136. index_input = str(input("What Index would you like to compare your portfolio to?\n"))
137. index_data = pdr.DataReader(index_input, 'yahoo', start_date, end_date)['Adj Close']
138.
139. #Function that calculates the starting price of one share of the portfolio given the row number.
140. def portfolio_calc(row):
141.     price = 0
142.     if index_input == "SPY":
143.         for column in range (0,len(tickers)):
144.             price += (data[tickers[column]].iloc[row]*WEIGHT_DICTIONARY[tickers[column]])
145.             sum_of_shares += WEIGHT_DICTIONARY[tickers[column]]
146.         price = price/sum_of_shares
147.     if index_input == "^DJI":
148.         price = data.sum(axis=1)[row]/len(tickers)
149.     if index_input != "^DJI" and "SPY":
150.         price = data.sum(axis=1)[row]
151.     return price
152.
153.
154. print("The starting price for your portfolio is ",portfolio_calc(0),".")
155.
156. index_list = []
157. portfolio_list = []
158.
159. for r in range (0,len(data.index)):
160.     index_list.append (index_data.iloc[r])
161.     portfolio_list.append(portfolio_calc(r))
162.
163. comparison = pd.DataFrame(index=data.index)
164. comparison['Index Price'] = index_list
165. comparison['Portfolio Price'] = portfolio_list
166.
167. #Makes a new DataFrame that changes dollar price to percent change from the value prior.
168. comparison_percentage = pd.DataFrame( index = data.index)
169.
170.
171. def percent_change(Column,row):
172.     if Column == 1:
173.         percentage = ((comparison['Index Price'].iloc[r] - comparison['Index Price'].iloc[0])
174.                      / comparison['Index Price'].iloc[0])
175.     if Column == 2:
176.         percentage = ((comparison['Portfolio Price'].iloc[r] - comparison['Portfolio Price'].iloc[0])
177.                      / comparison['Portfolio Price'].iloc[0])
178.     return percentage
179.
180.
181. index_list_percent = []
182. portfolio_price_percent = []
183.
184. for r in range (0,len(comparison_percentage.index)):
185.     index_list_percent.append(percent_change(1,r))
186.     portfolio_price_percent.append(percent_change(2,r))
187.
188. comparison_percentage['Index Price'] = index_list_percent
189. comparison_percentage['Portfolio Price'] = portfolio_price_percent
190.
191. outperformance_count = 0
192. total_count = 0
193.
194. for r in range (0, len(comparison_percentage.index)):
```

```
195.        if comparison_percentage['Portfolio Price'].iloc[r] > comparison_percentage['Index Price'].ilo
    c[r]:
196.            outperformance_count += 1
197.        total_count += 1
198. outperformance_percentage = (outperformance_count / total_count) * 100
199.
200. print(comparison_percentage)
201. print("Your portfolio, which consists of", tickers, ",", outperforms", index_input, outperformance_p
    ercentage,"% of the time.")
202.
203. comparison_percentage.plot()
204. plt.show()
```

## C.       Investing with Stock Indicators Program

```
1.   """
2.   Tests investing using indicators
3.
4.   Install:
5.   $ pip install fix_yahoo_finance --upgrade --no-cache-dir
6.   pip install pandas-datareader
7.   """
8.   from pandas_datareader import data as pdr
9.   import matplotlib.pyplot as plt
10.  import pandas as pd
11.  import numpy as np
12.  import fix_yahoo_finance as yf
13.  import networkx as nx
14.
15.
16.  yf.pdr_override()
17.
18.  indicators = []
19.
20.  ticker_input = input("Please enter the stock ticker you have chosen to invest in. \n")
21.
22.  indicator_input = input("Please enter the stock ticker you have chosen to be in your portfolio.Whe
     n you are done just type 'DONE'! \n")
23.
24.  while indicator_input != "DONE":
25.      indicators.append(str(indicator_input))
26.      indicator_input = input()
27.
28.  start_date= input("Here enter the date from which data will begin to be taken from (Please put it
     into format of YYYY-MM-DD)\n")
29.
30.  data_end_date= input("Here enter the date at which we will stop taking data from (Please put it in
     to format of YYYY-MM-DD)\n")
31.
32.  emulation_end_date = input("Here enter the date that the investing simulation will go up to (Pleas
     e put it into format of YYYY-MM-DD)\n")
33.
34.  #Downloads data from Yahoo using inputs given by the user.
35.  ticker_data = pdr.DataReader(ticker_input, 'yahoo', start_date, data_end_date)['Adj Close']
36.  indicator_data = pdr.DataReader(indicators, 'yahoo', start_date, data_end_date)['Adj Close']
37.
38.
39.  def corr(x,y,days):
40.      x_avg = x.mean()
```

```python
41.        y_avg = y.mean()
42.        numerator_sum = 0
43.        denominator_sum_x = 0
44.        denominator_sum_y = 0
45.        for i in range(days,len(x)):
46.            numerator_sum += (x[i] - x_avg)*(y[i-days] - y_avg)
47.            denominator_sum_x += (x[i] - x_avg) ** 2
48.            denominator_sum_y += (y[i-days] - y_avg) ** 2
49.        correlation_value = (numerator_sum / ((denominator_sum_x*denominator_sum_y) ** (0.5)))
50.        return correlation_value
51.
52. OPTIMAL_SHIFT_DICTIONARY = {}
53. OPTIMAL_CORR_DICTIONARY = {}
54.
55. ticker_emulation_data = pdr.DataReader(ticker_input, 'yahoo', data_end_date, emulation_end_date)['Adj Close']
56. indicator_emulation_data = pdr.DataReader(indicators, 'yahoo', data_end_date, emulation_end_date)['Adj Close']
57.
58. for indicator_index in range (0,len(indicators)):
59.     highest_correlation = corr(ticker_data, indicator_data[indicators[indicator_index]],1)
60.     highest_days = 1
61.     for n in range (1,80):
62.         if corr(ticker_data, indicator_data[indicators[indicator_index]],n) > highest_correlation:
63.             highest_correlation = corr(ticker_data, indicator_data[indicators[indicator_index]],n)
64.         if n > highest_days:
65.             highest_days = n
66.
67.     OPTIMAL_SHIFT_DICTIONARY[indicators[indicator_index]] = highest_days
68.     OPTIMAL_CORR_DICTIONARY[indicators[indicator_index]] = highest_correlation
69. print(OPTIMAL_SHIFT_DICTIONARY)
70. print(OPTIMAL_CORR_DICTIONARY)
71.
72. ticker_emulation_data_pct = ticker_emulation_data.pct_change()
73. indicator_emulation_data_pct = indicator_emulation_data.pct_change()
74.
75.
76. def test(r):
77.     conditionals = 0
78.     for indicator_index in range (0,len(indicators)):
79.         if indicator_emulation_data_pct[indicators[indicator_index]][r-OPTIMAL_SHIFT_DICTIONARY[indicators[indicator_index]]] > 0:
80.             conditionals += 1
81.     if conditionals >= number_of_true:
82.         return "true"
83.     else:
84.         return "false"
85.
86.
87. portfolio_price = ticker_emulation_data[0]
88.
89. print(len(indicators),"relationships have been developed. How many of these relationships must be true in order to invest in"
90.       ,ticker_input,"To try a new number close window and type in a new number. If you are done enter any negative number.")
91. number_of_true = int(input())
92.
93. while number_of_true >= 0:
94.     continuous_investing = ticker_emulation_data[:]
```

```
95.        indicative_investing = []
96.        for r in range (0,len(ticker_emulation_data.index)):
97.            if number_of_true == 0:
98.                if r == 0:
99.                    last_true = portfolio_price
100.                if r != 0:
101.                    last_true = (last_true+ (ticker_emulation_data[r]-ticker_emulation_data[r-1]))
102.            else:
103.                if r < highest_days:
104.                    last_true = portfolio_price
105.                if r >= highest_days:
106.                    if test(r) == "true":
107.                        last_true = (last_true+ (ticker_emulation_data[r]-ticker_emulation_data[r-1]))
108.            indicative_investing.append(last_true)
109.
110.        comparison = pd.DataFrame(index = ticker_emulation_data.index)
111.        comparison['Continuous Investing'] = continuous_investing
112.        comparison['Indicative Investing'] = indicative_investing
113.
114.        print(comparison)
115.
116.        comparison.plot()
117.        plt.show()
118.        number_of_true = int(input())
119.
120. labels={}
121. labels[0] = ticker_input
122. for i in range (1,len(indicators)):
123.     labels[i] = indicators[i]
124.
125. graph = {}
126.
127. for indicator_index in range (0,len(indicators)):
128.     if indicator_emulation_data_pct[indicators[indicator_index]][(len(ticker_emulation_data.index))-OPTIMAL_SHIFT_DICTIONARY[indicators[indicator_index]]] > 0:
129.         graph[indicators[indicator_index]] = '1'
130.
131.
132. labels = {}
133. labels['1'] = ticker_input
134.
135. print(graph)
136. G = nx.DiGraph(graph)
137. G.add_nodes_from(indicators)
138. H = nx.relabel_nodes(G,labels)
139. nx.draw(H,node_color='black',with_labels=True,node_size=800,font_size=10,font_color="white")
140. plt.show()
```